\documentclass[]{jfm}

\usepackage[FIGTOPCAP]{subfigure} 

\usepackage{graphicx}
\usepackage{psfrag}
\usepackage{braket}
\usepackage{natbib}
\usepackage{color}
\definecolor{grey}{rgb}{0.5,0.5,0.5}

\ifCUPmtlplainloaded \else
  \checkfont{eurm10}
  \iffontfound
    \IfFileExists{upmath.sty}
      {\typeout{^^JFound AMS Euler Roman fonts on the system,
                   using the 'upmath' package.^^J}%
       \usepackage{upmath}}
      {\typeout{^^JFound AMS Euler Roman fonts on the system, but you
                   dont seem to have the}%
       \typeout{'upmath' package installed. JFM.cls can take advantage
                 of these fonts,^^Jif you use 'upmath' package.^^J}%
      }
  \else
  \fi
\fi

\ifCUPmtlplainloaded \else
  \checkfont{msam10}
  \iffontfound
    \IfFileExists{amssymb.sty}
      {\typeout{^^JFound AMS Symbol fonts on the system, using the
                'amssymb' package.^^J}%
       \usepackage{amssymb}%
       \let\le=\leqslant  \let\leq=\leqslant
         
      }{}
  \fi
\fi

\ifCUPmtlplainloaded \else
  \IfFileExists{amsbsy.sty}
    {\typeout{^^JFound the 'amsbsy' package on the system, using it.^^J}%
     \usepackage{amsbsy}}
    {\providecommand\boldsymbol[1]{\mbox{\boldmath $##1$}}}
\fi

\newcommand\Rey{\mbox{\textit{Re}}}  
\newcommand\De{\mbox{\textit{De}}}  

\newsavebox{\astrutbox}
\sbox{\astrutbox}{\rule[-5pt]{0pt}{20pt}}

\title{Experimental investigation of transitional flow in a toroidal pipe}

\author[J.\ K\"{u}hnen, M.\ Holzner, B.\ Hof and H.\ C.\ Kuhlmann]%
{J.\ns K\ls \"{U}\ls H\ls N\ls E\ls N\ls $^1$%
  \thanks{Email address for correspondence: jakob.kuehnen@ist.ac.at},\ns
M.\ns H\ls O\ls L\ls Z\ls N\ls E\ls R\ls $^2$,
B.\ns H\ls O\ls F\ls $^3$, 
\and H.\ns C.\ns K\ls U\ls H\ls L\ls M\ls A\ls N\ls N$^1$}

\affiliation{
$^1$Institute of Fluid Mechanics and Heat Transfer, Vienna University of Technology,
Resselgasse 3/1/2, A-1040 Vienna\\[\affilskip]
$^2$Institute of Environmental Engineering, ETH Z\"{u}rich, Wolfgang Pauli Str.\ 15, CH-8093
Z\"{u}rich\\[\affilskip]
$^3$Max Planck Institute for Dynamics and Self Organization, Bunsenstrasse 10, D-37073
G\"{o}ttingen}

\pubyear{2012}
\volume{650}
\pagerange{119--126}
\date{?; revised ?; accepted ?. - To be entered by editorial office}

\graphicspath{{./figures/}}

\begin{document}

\maketitle

\begin{abstract}
The flow instability and further transition to turbulence in a toroidal
pipe (torus) with curvature (tube-to-coiling diameter) 0.049 is investigated
experimentally. The flow inside the toroidal pipe is driven by a steel sphere fitted to
the inner pipe diameter. The sphere is moved with constant azimuthal velocity from
outside the torus by a moving magnet. The experiment is designed to investigate curved
pipe flow by optical measurement techniques. Using stereoscopic particle image
velocimetry, laser Doppler velocimetry and pressure drop measurements the flow is
measured for Reynolds numbers ranging from 1000 to 15\,000. Time- and space-resolved
velocity fields are obtained and analyzed. The steady axisymmetric basic flow is strongly
influenced by centrifugal effects. On an increase of the Reynolds number we find a
sequence of bifurcations. For $\Rey=4075\pm2\%$ a supercritical bifurcation to an oscillatory
flow is found in which waves travel in streamwise direction with a phase velocity
slightly faster than the mean flow. The oscillatory flow is superseded by a presumably
quasi-periodic flow at a further increase of the Reynolds number before turbulence sets
in. The results are found to be compatible, in general, with earlier experimental and
numerical investigations on transition to turbulence in helical and curved pipes.
However, important aspects of the bifurcation scenario differ considerably.
\end{abstract}



\section{Introduction}\label{sec:intro}

Transition to turbulence in \textit{straight} circular pipes is one of the oldest and
most fundamental problems of fluid mechanics, entailing various fundamental questions
about the nature of turbulence \citep{Eckhardt2008}. The understanding of
the physics of transition to turbulence in straight pipes has experienced a significant
progress in recent years due to the application of modern optical measurement techniques
and computationally-based theoretical modeling \citep{Hof2004,Eckhardt2007,Mullin2011}.
It is characteristic for straight pipe flow that transition occurs despite of the linear
stability of the laminar Hagen--Poiseuille flow if the flow perturbations exceed a
certain threshold. Moreover, and different from plane Poiseuille flow, Rayleigh–-B\'{e}nard
convection, or Taylor--Couette flow, there exists no instability of Hagen--Poiseuille flow for any Reynolds number $\Rey= U d/{\nu}$ where $U$ is the mean
velocity, $d$ the diameter of the tube and $\nu$ the kinematic viscosity  of the fluid
\citep{Drazin1981}. Consequently, turbulence can only be triggered by finite amplitude perturbations. Following \cite{Reynolds1883} the critical point is then defined as the Reynolds number above which turbulence is first sustained indefinitely, whereas below any turbulence initially in the flow will eventually decay. As shown by \cite{Avila2011} this critical Reynolds number is $\Rey_c \approx2040$ where the decay of turbulence is balanced by a spreading process.


In \textit{curved} pipes the process of transition to turbulence qualitatively differs from that in straight pipes.
Curved-pipe flow has received less attention than straight-pipe flow, even though the
former is of no lesser importance: practically all pipes in nature, e.g.\ blood vessels,
or in engineering are curved. In curved pipes the basic flow is strongly affected by an
imbalance between the cross-stream pressure gradient and the centrifugal force which
leads to a secondary cross-stream motion, typically in form of a pair of steady
streamwise Dean vortices symmetric with respect to the tangent plane. With increasing
Dean number the maximum of the streamwise velocity is shifted radially towards the outer
wall of the pipe. Owing to the increased gradient of the streamwise flow the drag in
curved pipes is considerably higher than in straight pipes. For reviews of curved-pipe
flow and further reading we refer to \cite{Berger1983,Huttl2000,
Huttl2001,Naphon2006,Vashisth2008}.

\cite{Dean1927,Dean1928} has solved the simplified Navier--Stokes equations for a coiled
pipe of small curvature showing that the flow is governed by two parameters. These
parameters are the curvature $\delta=d/D$ of the pipe diameter $d$ to the coiling
diameter $D$ and the Dean number $\De=\Rey\sqrt{\delta}$. \cite{Taylor1929},
\cite{White1929} and \cite{Adler1934} found that the flow in curved pipes remains laminar
up to Reynolds numbers higher at least by a factor of two than in straight pipes. They
also noticed that transition to turbulence is not as abrupt as in straight pipes but
occurs gradually, without any discontinuity of the characteristic observables such as the
pressure drop. \cite{Sreenivasan1983} investigated curved-pipe flow for moderate Reynolds
numbers. They found that the turbulent flow in a straight pipe becomes laminar after
entering a coiled-pipe of the same diameter. They recognized, moreover, that the
curved-pipe flow becomes time periodic before turbulence commences. The oscillation
amplitude was found to be strongest near the inside wall of the pipe.

\cite{Webster1993} investigated the unsteady three-dimensional flow through a helical
pipe at transitional conditions using laser-Doppler velocimetry. For $\delta=0.055$ they
measured the onset of periodic low-frequency perturbation waves at $\Rey_c = 5060$. They
specified the nondimensional frequencies found with $fd/U=0.25$ and 0.5 for all Reynolds numbers investigated. Complementary numerical simulations for $\Rey=5480$ by \cite{Webster1997} were in agreement with the experimental results of \cite{Webster1993}. Dye streaks, used for visualization, were found to diffuse at $\Rey=6330$. \cite{Webster1997} interpreted this behavior as the onset of turbulent fluctuations.

Flows in weakly curved ducts with constant curvature and in helical pipes
with small torsion have a common asymptotic limit: toroidal-pipe flow. However, the inlet
and outlet conditions are generally different. \cite{Piazza2011} numerically simulated
the flow in a circular toroid with a circular cross-section which is driven by a
constant streamwise body force, using periodic boundary conditions.
For $\delta=0.3$ a supercritical Hopf bifurcation was found in the
interval $4556<\Rey_{c1}<4605$, giving rise to an azimuthally traveling wave which took the form of a
varicose streamwise modulation of the two Dean vortex rings. The wave was described to
mainly affect the Dean vortices but not the boundary layers on the wall. For
$5042<\Rey_{c2}<5270$ a secondary Hopf bifurcation was discovered leading to a
quasi-periodic flow. The second wave, created for $\Rey>\Rey_{c2}$, arises as an array of
oblique vortices localized in the two boundary layers of the basic flow at the edge of
the Dean vortex regions. The oscillatory perturbation flows in both the periodic and the
quasi-periodic regimes were anti-symmetric with respect to the equatorial midplane of the
torus for $\delta=0.3$. For $\delta=0.1$ \cite{Piazza2011} found a direct transition from
a steady to a quasi-periodic flow between $5139<\Rey<5208$, associated with hysteresis.
The traveling waves for $\delta=0.1$ were symmetric with respect to the equatorial
midplane of the torus.

Like \cite{Cioncolini2006}, most studies on curved-pipe flow aimed at
reliable pressure-drop correlations, because the friction factor is an important quantity
for industrial design. Moreover, all experimental studies on transition to turbulence in
fully developed curved pipes have been conducted in helical pipes with a small pitch.
While the closed flow in a torus considered by \cite{Piazza2011} offers a perfect geometry for  theoretical and numerical
investigations, experiments using a torus are difficult, because the driving pressure
gradient cannot as easily be imposed as in open systems. Apart from \cite{Madden1994},
\cite{Pino2008} and \cite{Hewitt2011}, who experimentally studied the \textit{transient}
flow in a torus during spin-up from rest and spin-down from solid-body rotation, we are
not aware of any other experimental work on the flow in a torus.

In the present work we experimentally investigate the flow and its instability in a torus
by means of visual observations, high-speed stereoscopic particle image velocimetry
(SPIV) and pressure drop measurements. Particular attention is paid to the detection and
investigation of traveling waves. To that end an experiment was set up where the flow inside a torus is driven by a
moving sphere. For a sufficiently large aspect ratio (circumference-to-pipe-diameter ratio) the perturbations induced by the sphere remain localized and bulk flow properties can be measured which are independent of the particular driving.
We measure the full three-dimensional and time-resolved structure of
the transitional flow including all three velocity-components over the entire length and
cross-section of the pipe.

\section{Experimental setup}\label{sec:experimenalsetup}

\subsection{The facility}\label{subsec:facility}

\begin{figure}
  \centerline{\includegraphics[clip,width=0.72\textwidth,angle=0]{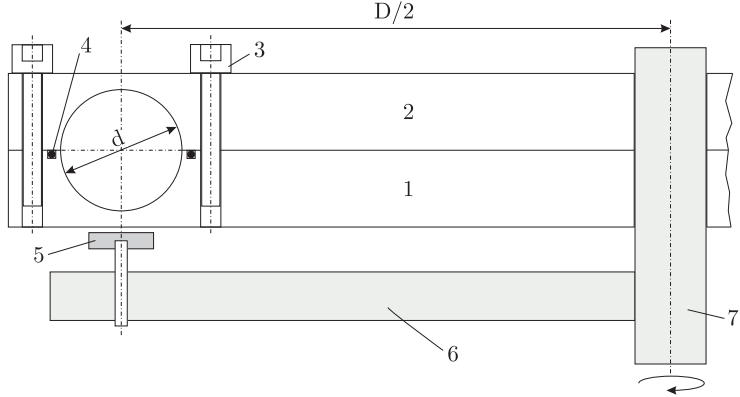}}
  \caption{\label{fig:Torus-Sketch}Assembly drawing (side view) of the experimental setup.
  Lower (1) and upper (2) plexiglass disk are bolted (3) and sealed using a
  2.5\,mm rubber O-ring (4) in a 2.2\, x 2.5\,mm notch. A total of 32 (16 outside and 16
  within the torus-notch, in an angular distance of $\pi/8$) screws is used to ensure leakproof
  tightness. A flat pot magnet (5) with threaded stem is incorporated to the boom (6) right
  below the centerline of the tube, leaving an adjustable distance of 1\,mm to the lower
  plexiglass disk. The boom is rotated around the shaft (7) by a geared direct current motor (not shown).
  The diameter of the torus is $D=614$\,mm, the diameter of the tube is $d=30.3$\,mm. Drawing not to scale. }
\end{figure}

The toroidal pipe is realized as a toroidal cavity (torus) in a
stationary block of plexiglas. A ferromagnetic sphere,  which is actuated by a strong
magnet from the outside, is put into the toroidal cavity to drive the flow in the
cavity. The main components and the driving mechanism of the experiment are sketched in
figure \ref{fig:Torus-Sketch}. The stationary block of plexiglas is
built of two disks of highly transparent and polished perspex into which a concentric
notch of semi-circular cross sections had been machined. The two disks are mounted
mirror-symmetrically. The two notches in the disks hence form a closed toroidal cavity,
i.e.\ a curved tube which is to be filled with the working fluid. To drive the fluid
motion a ferromagnetic stainless chromium steel sphere with diameter slightly less than
the toroidal tube is placed in the toroid. The steel sphere is actuated from outside the
toroidal cavity using a strong permanent magnet mounted on a rotating boom. To achieve a
constant and precisely adjustable flow rate in the torus the boom is rotated at a
constant angular velocity, thereby steadily moving the sphere inside the torus, driving
the fluid motion. The driving system to rotate the shaft consists of an electric gear
motor combined with a belt drive (both not shown in figure \ref{fig:Torus-Sketch}). The
Reynolds number is defined as $\Rey=U d/{\nu}$, where $d$ is the diameter of the tube,
$\nu$ the kinematic viscosity of the fluid, and the bulk velocity $U=\Omega D/2$ with
angular velocity $\Omega$ of the rotating boom and $D$ the diameter of the centre circle
of the torus.

When measuring flow details with SPIV a large tube diameter
$d$ is desirable. Therefore, a tube diameter of $d=30.3$\,mm was chosen, slightly larger
than the diameter of the sphere $d_{s}=30$\,mm, to permit rolling motion of the sphere.
This left a small sickle-shaped gap of maximum $0.3$\,mm at the upper half between torus
wall and sphere. The total area of the gap is $A_{gap}=14.2$\,mm$^{2}$ corresponding to
1.9\% of the tube's cross section. Through the gap between sphere and tube wall a small
leak flow will arise. When the fluid in the torus has reached a constant velocity after
spin up from rest the sphere has to balance just the minor friction losses in the tube
which amounted to, e.g.\ $\approx20$\,Pa at $\Rey=5000$.

Being interested in a small curvature $\delta$ a large diameter $D$ of the torus is
preferable. We used a centerline diameter of $D=614$\,mm, resulting in a curvature $\delta=d/D=0.049$. This corresponds to a large aspect ratio (circumference to tube diameter) of $\Gamma=\pi/\delta - 1 \approx 63$. To achieve a surface finish of the tube comparable to that of glass
pipes the perspex surface was polished. The torus dimensions including the manufacturing
tolerances are $D=614\pm0.1$\,mm and $d=30.3\pm0.03$\,mm, respectively.

According to the manufacturer's specification the pot magnet had $\approx 216$\,N
adhesive force perpendicular to the surface to which the magnet should stick (equivalent
to 22\,kg bearing capacity). However, the actual adhesive force decreases quickly with
increasing distance and is also influenced by the alloy of the object of magnetic
attraction. Since the minimum distance between magnet and sphere in the setup is
$\approx5$\,mm (see fig.\ref{fig:Torus-Sketch}), the actual adhesive force imposed on the
sphere is about 0.49 to 0.98\,N (0.05 to 0.1\,kg) according to specification. This was
found sufficient to control the sphere in the liquid-filled torus. Owing to
the high inertia of the rolling sphere compared to any forces caused by the flow
fluctuations the sphere would always roll very smoothly and accurately follow the magnet.
Observations of the sphere by a camera mounted on the boom did not allow to detect any
relative motion between the sphere and the magnet, once the steady rolling motion was
established. Only in the case of very fast or jerky acceleration or deceleration the
sphere would not follow the magnet and loose the magnetic bond.

To detect the temperature of the fluid in the torus, two small Platinum SMD Flat Chip
temperature sensors (\textit{Vishay Beyschlag}, PTS 0603, $100\,\Omega$) are used. Every
sensor is mounted on the front end of a bracket pin with a diameter of 2\,mm. The bracket
pins are plugged in holes which were drilled into the upper plexiglas disk in an angular
distance of $\pi/2$, just above the upper apex of the tube. The holes would just not
shove through into the toroidal volume but leave a thin indentation of approximately
0.08\,mm between the temperature sensors and the fluid in the torus. For data acquisition
the two sensors are connected to a \textit{Validyne} UPC 2100 PCI sensor interface card.
The mean value of the two sensor signals, averaged over 20\,s, is used to determine the
temperature of the fluid in the torus with an accuracy of $\pm0.1$\,K.
The ambient temperature in the laboratory could vary slowly by
approximately $\pm2$\,K during a day. As individual measurements would never take longer
than 5 minutes no additional measures were taken to control the temperature of the
fluid. To eliminate problems due to the differing refractive indices such as unwanted
displacement, hidden regions and multiple images \citep{Lowe1992}, arising for objects
placed inside containers with cylindrical walls, a refractive-index-matched mixture of
distilled water and ammonium thiocyanate was used as working fluid \citep{Budwig1994,
Hopkins2000}. The kinematic viscosity $\nu$ of the index-matched fluid as a function of
temperature was determined by means of a \textit{Schott} Cannon-Fenske capillary
viscometer. For temperatures ranging from 20 to 30\,$^{\circ}$C the kinematic viscosity
of the index-matched fluid was approximated by the fourth-order polynomial
\begin{equation}\label{eq:visc-matched}
    \nu= 4.5\times10^{-7}+2.7\times10^{-7}\, T-1.8\times10^{-8}\, T^{2}+4.9\times10^{-10}\, T^{3}-4.9\times 10^{-12}\, T^{4},
\end{equation}
which was fitted to the measured data by least squares. Here, $T$ is measured in degrees
Celsius and $\nu$ in m$^2$/s. A combination of a \textit{LabView}-program and a system
consisting of controller unit (LSC 30/2, 4-Q-DC, \textit{maxon motor ag}), USB data
acquisition card (\textit{National Instruments} 6008) and direct current motor (RE 25,
\textit{maxon motor ag}) with speedometer (DCT 22) and planetary gear (GP 32) was used to
set and continually control the Reynolds number. The
\textit{LabView}-program contained the functional relationship between temperature,
viscosity and the geometrical data based on which the required angular velocity of the
sphere for a specified Reynolds number was calculated and implemented. Optionally, an
automatic mode can be used to increase or decrease the Reynolds number stepwise by
predefined step sizes and at predefined times. The setup provided the possibility to
investigate the flow in the torus for adjustable Reynolds numbers ranging from 1\,000 to
15\,000 with an accuracy of $\pm 2$\% taking into account the leak flow past the sphere (see \ref{subsubsec:leakflow}).

\subsection{Measurements}\label{subsec:measurements}

For preliminary investigations and visual identification of pertinent flow structures,
the fluid was seeded with glitter particles (polyester glitter, \textit{Sigmund Lindner
GmbH}) which reflect the light and visualize the flow structure. However, the conspicuity
was limited, because the particles tended to get stuck in the gap between sphere and
torus wall and would hence block the sphere if their concentration was too high.
Therefore, only a limited amount of particles could be used. Still photographs as well as
video recordings were made with a \textit{Nikon} D 7000 DSLR camera. Movies of the flow
structures were recorded with a frame size of $1920\times 1080$ pixels at 24\,fps. For
the video recordings the camera was mounted on an additional boom which rotated at the
same speed as the boom driving the sphere. This allowed observations in the frame of
reference rotating with the angular velocity of the sphere which nearly equals the mean
flow velocity $U$.

The pressure drop between two pressure holes (bore holes with internal diameter of
1\,mm) was determined using a differential pressure sensor (\textit{Validyne} DP103,
ultra low range wet-wet differential pressure transducer). The pressure transducer was
calibrated using a Betz manometer and it had a full range of 140\,Pa with an accuracy
better than 0.35\,Pa. The pressure holes were located at the top of the upper side of
the tube. The sampling rate was 10\,ms.

\begin{figure}
  \centerline{\includegraphics[clip,width=0.66\textwidth,angle=0]{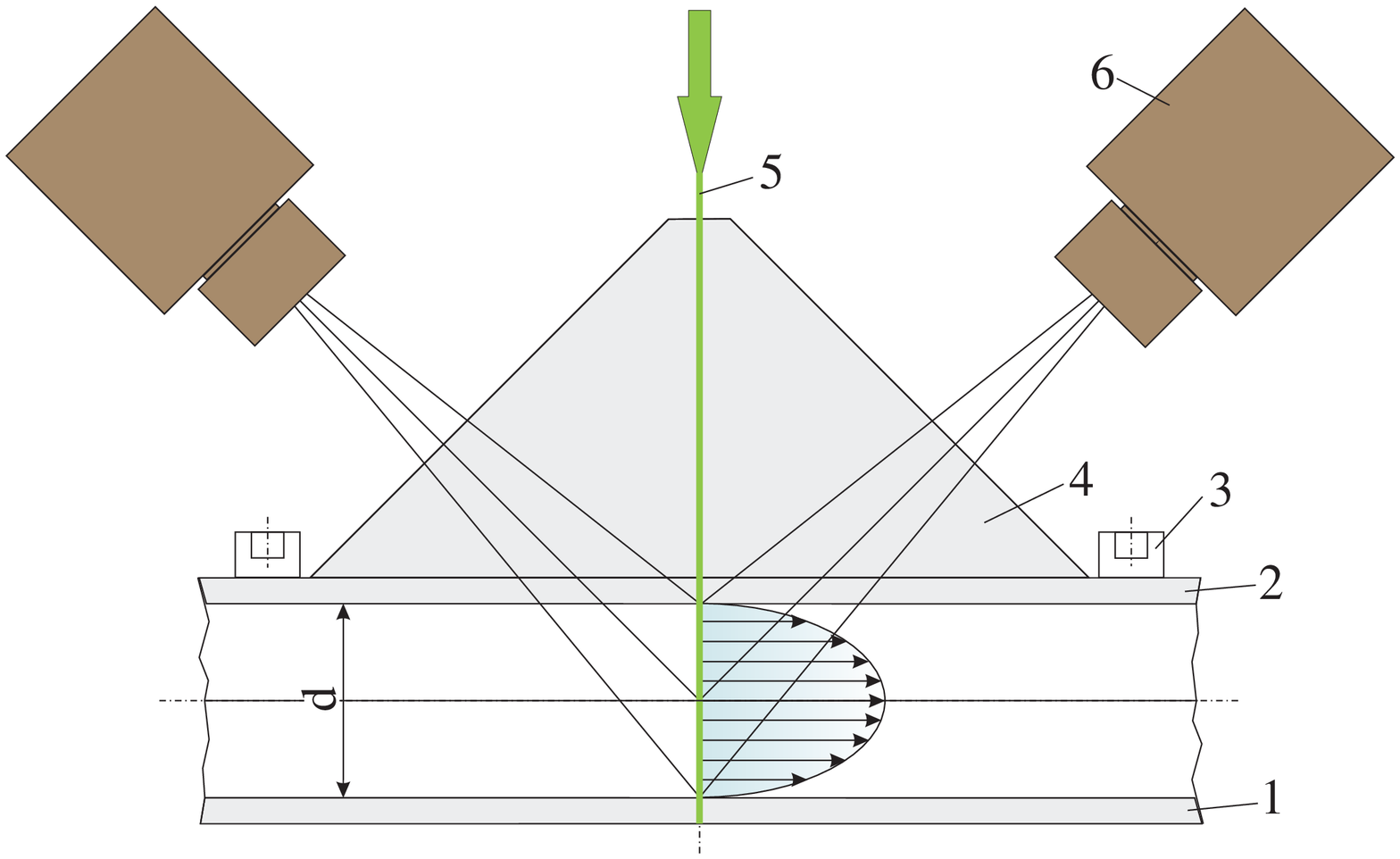}}
  \caption{\label{fig:PrismaAufbau-Sketch} Sketch (side view) of the stereoscopic PIV
  system. To reduce optical reflections, a prism is placed on the upper plexiglas disk of
  the torus. (1) lower plexiglass disk, (2) upper plexiglass disk, (3) screws, (4) prism, (5)
  light sheet, (6) cameras. A few drops of glycerine were used to create a thin film between (2) and (4), enabling seamless
  contact. Drawing not to scale.}
\end{figure}

A SPIV system from \textit{LaVision} has been used to measure
the three components of the velocity vectors over a cross section perpendicular to the
flow. It consisted of a pulsed \textit{Quantronix} Darwin Duo laser (diode pumped Nd:YLF
laser, wavelength 527\,nm, 60\,mJ total energy) and two Phantom V10
high-speed cameras with a full resolution of $2400\times 1900$\,px. The temporal
resolution was set to 100\,Hz. Silver-coated hollow glass spheres (S-HGS-10, mean
diameter 10\,$\mu$m, $\rho=1.4$\,g/cm$^3$, \textit{Dantec Dynamics}) were used as
seeding particles.

Previous investigators \citep[see e.g.][]{Doorne07} have demonstrated the possibilities
and potential of SPIV to investigate and capture the appearance and development of
transitional flow in a pipe. The same principle is applied in the experimental setup used
for this work. Figure \ref{fig:PrismaAufbau-Sketch} provides a sketch of the setup which
was used for stereoscopic PIV, showing the arrangement of plexiglass disks, light sheet
and 2 cameras in a 45$^\circ$ viewing angle. Due to the 45$^\circ$ viewing angle the
light would be refracted at the outer wall of the plexiglass plate, which would reduce
the effective viewing angle of the cameras and degrade the image quality. Therefore, a
prism made of perspex is attached to the upper plexiglass plate such that the optical
axis is perpendicular to air--perspex interface.

The calibration target, needed for the calibration of the SPIV measurements, was custom
made by \textit{Die Signmaker GmbH} (G\"{o}ttingen). It consisted of 5\,mm--spaced lattice of
black dots with a diameter of 1\,mm printed on both sides of a 1.5\,mm perspex disk with
a cross-sectional diameter of 30\,mm, i.e.\ slightly less than the tube diameter. The
disk was kept in position (congruent with the light sheet of the laser) for the
calibration procedure by a tiny locking screw drilled into the torus wall. After the
calibration images were shot the screw was removed and the calibration target was
released. The experimental setup was too sensitive to be disassembled after calibration,
as dismantling and repeated assembly would have caused small displacements and hence
spoiled the calibration. Therefore, the disk was just left in the torus and
went with the flow. After a few turns of the actuator the disk would then be
located in front of the sphere, not perturbing or changing the flow far from the sphere
in addition to the perturbations induced by the actuator itself.

The evaluation of the 3D-vector fields from the PIV images was performed with commercial
PIV-software (DaVis 8, \textit{LaVision}). The interrogation area was $32\times
32$\,px$^2$ with an overlap of 50\%. The maximum particle displacement between two frames
was approximately 11\,px in horizontal and 8\,px in vertical direction. Subsequently, the
acquired data were analyzed with \textit{Mathworks} Matlab R2011b which
was used as programming environment.

A fibre-flow 2D-LDV system of \textit{Dantec Dynamics} was employed for
laser-Doppler velocimetry measurements. From the LDV data we obtained spectra of the streamwise
velocity at single points within the cross section of the pipe. The average sampling rate
was 200\,Hz.

\section{Results}\label{sec:results}

\begin{figure}
\psfrag{x}[bc][bc]{$x$}
\psfrag{y}[bc][bc]{$y$}
\psfrag{phi}[bl][bl]{$\alpha$}
\psfrag{A}[bc][bc]{A}
\psfrag{B}[bc][bc]{B}
\psfrag{C}[bc][bc]{C}
\psfrag{D}[bc][bc]{D}
\psfrag{hs}[bc][bc]{hs}
\psfrag{ls}[bc][bc]{ls}
\psfrag{core}[br][br]{core}
\psfrag{region I}[bc][bc]{region I}
\psfrag{CSWL}[br][br]{\parbox{1.8cm}{\centering cross-stream\\ wall layer}}
\psfrag{oec}[tr][tr]{outer equatorial point}
\psfrag{iec}[tl][tl]{inner equatorial point}
\centerline{\includegraphics[clip,width=0.38\textwidth,angle=0]{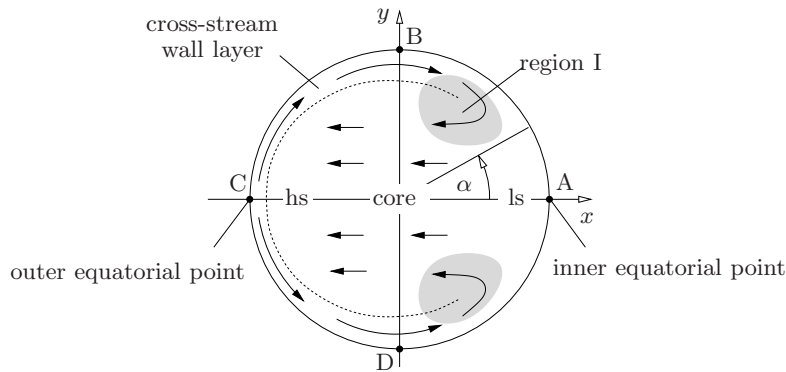}}
  \caption{\label{fig:Referenzsystem_figures}Salient flow structures in a cross section of the
  toroidal pipe and notation. A cartesian ($x,y,z$) coordinate system is used in the meridional
  observation plane at constant toroidal angle $\varphi$. The poloidal (meridional)
  angle is $\alpha$. The $x$-axis is directed toward the
  symmetry axis of the torus. The symbols hs and ls denote regions of high-speed
  and low-speed streamwise velocity, respectively. A,B,C and D denote the
  north-pole (B), south-pole (D), outer equatorial (C) and inner equatorial (A) points.}
\end{figure}

Most of the space-resolved measurements are taken in a meridional cross section of the
torus. Therefore, we use local Cartesian coordinates ($x,y,z$) as shown in figure
\ref{fig:Referenzsystem_figures}, where $(u,v,w)$ are the respective Cartesian velocity
components. The $x$-axis is directed radially inward towards the center of the torus.
Figure \ref{fig:Referenzsystem_figures} also sketches characteristic cross-stream flow
structures of the basic steady flow, where we use the same notation as
\cite{Webster1997}. The steady basic flow is mirror symmetric with respect to the equatorial
plane $y=0$. In a wide \textit{core region} around the axis $y=0$ the cross-stream flow
is directed radially outward in negative $x$-direction from the
\textit{low-speed core region} (ls) to the \textit{high-speed core region} (hs) of the streamwise velocity.
The cross-stream flow returns radially inward in two symmetrically located wall jets called
\textit{cross-stream wall layers}. The thickness of these wall jets grows as they turn
inward. Before reaching the inner equatorial point C the jets turn sharply in
\textit{region I} and merge to form the core flow in the low-speed region directed
radially outward.

\subsection{Evaluation of the experimental setup}\label{subsec:evaluationofexpsetup}

We are interested in the toroidal pipe flow driven by a prescribed steady
volume flux. Due to the particular driving by a moving sphere the ideal flow is
perturbed in the vicinity of the sphere. To assess to which extent the flow field in the torus can be considered
fully developed, i.e.\ stationary and independent of the driving mechanism, the effect of
the rolling sphere on the flow has to be evaluated. The perturbations arise in form of an
entrance-length effect and a leakage through the gap between the sphere and the torus. As
the curvature is constant throughout this work, the strength of the flow is
measured using the Reynolds number instead of the Dean number.

\subsubsection{Entrance-length effect}\label{subsubsec:influenceofactuator}

\begin{figure}
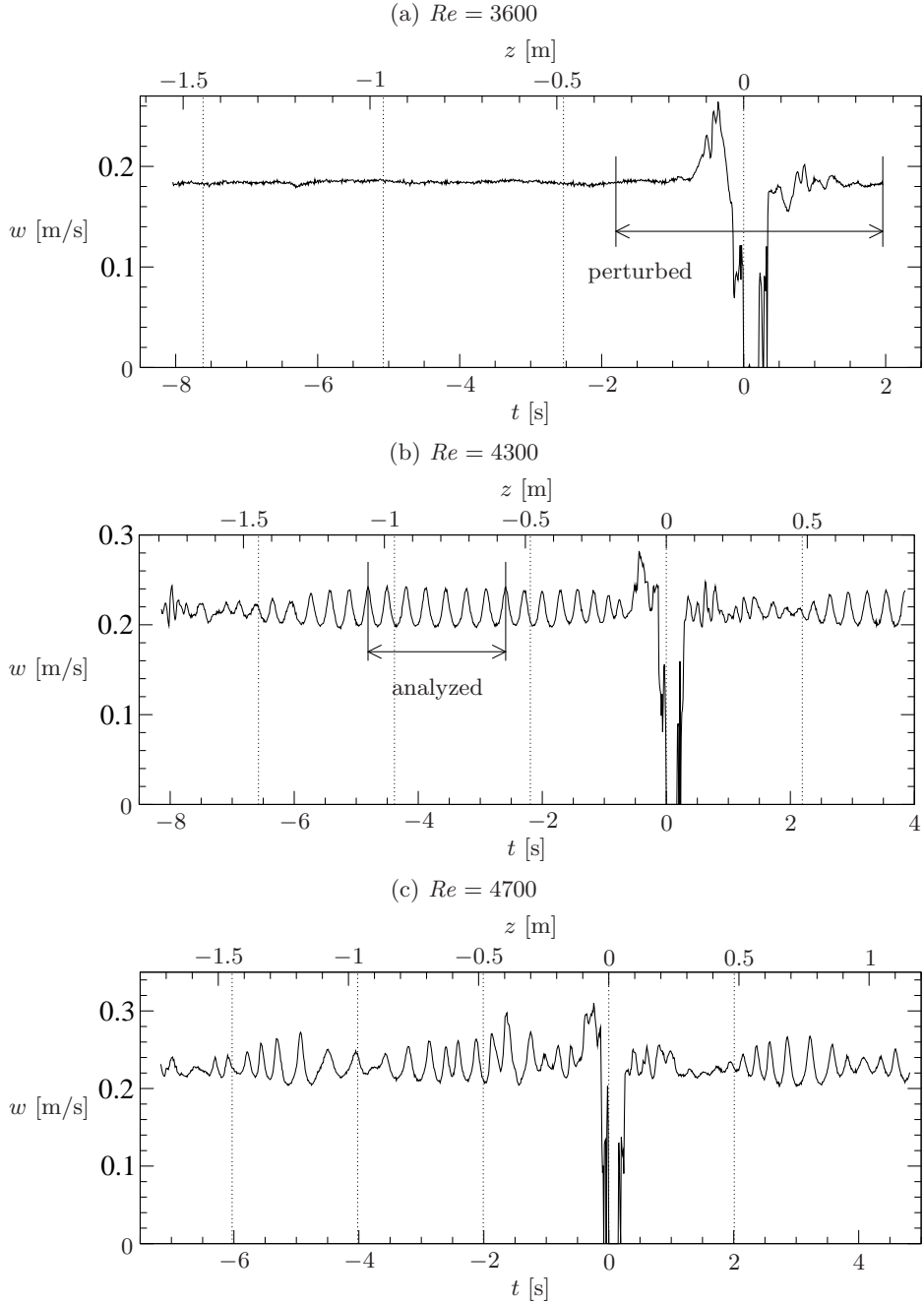

   \psfrag{w [m/s]}[bc][bc]{$w$ [m/s]}
   \psfrag{distance [m]}[bc][bc]{$z$ [m]}
   \psfrag{time [s]}[bc][bc]{$t$ [s]}
   \psfrag{-1.5}[bc][bc]{$-1.5$}
   \psfrag{-1}[bc][bc]{$-1$}
   \psfrag{-0.5}[bc][bc]{$-0.5$}
   \psfrag{0}[bc][bc]{$0$}
   \psfrag{0.5}[bc][bc]{$0.5$}
   \psfrag{1}[bc][bc]{$1$}
   \psfrag{-8}[bc][bc]{$-8$}
   \psfrag{-6}[bc][bc]{$-6$}
   \psfrag{-4}[bc][bc]{$-4$}
   \psfrag{-2}[bc][bc]{$-2$}
   \psfrag{2}[bc][bc]{$2$}
   \psfrag{4}[bc][bc]{$4$}
   \psfrag{perturbed}[bc][bc]{perturbed}
   \psfrag{unperturbed}[bc][bc]{analyzed}
\subfigure[$\Rey=3600$]{\includegraphics[clip,width=0.92\textwidth,angle=0]{3600.eps}}
\subfigure[$\Rey=4300$]{\includegraphics[clip,width=0.92\textwidth,angle=0]{4300.eps}}
\subfigure[$\Rey=4700$]{\includegraphics[clip,width=0.92\textwidth,angle=0]{4700.eps}}
\caption{\label{fig:onefullrevolution}Time traces of the streamwise velocity $w$ at $(x,y)=(0.17\,d,0.31\,d)$
for $\Rey=3600$, $\Rey=4300$ and $\Rey=4700$ as function of time $t$ (bottom axis) and azimuthal length $z$ (top axis).
A typical PIV-run of 12\,s, comprising approximately one full revolution of the sphere, is displayed.
The coordinate origin was defined by the transit of the sphere. The vertical dotted lines are drawn with a
spacing $\Delta\varphi=\pi/2$. The velocities of the sphere are $U_s=0.19$ (a), 0.22 (b) and 0.24 [m/s] (c).} \end{figure}

If the sphere moved as a plug with constant velocity without rolling a
pure streamwise and steady velocity field would be enforced at the surface of the sphere.
Due to the imposed velocity on the semi-spherical end wall and the velocity discontinuity
along the line of contact, there exists a return flow in the frame of reference moving
with the sphere which represents a perturbation to the fully developed toroidal-pipe flow
decaying away from the sphere. In operation the sphere is rolling in the torus.
While the rolling motion prevents any perturbation of stick--slip type and thus adds to
keeping the translational velocity constant, provided the rotation rate of
the beam is constant, the rolling motion induces an additional
cross-stream velocity component which will also decay away from the sphere
under subcritical conditions.

To quantify the decay of the above perturbations we measured the streamwise velocity $w$
as function of the distance from the sphere using SPIV. Figure
\ref{fig:onefullrevolution} shows the result as function of time (related to the distance
from the sphere by $\bar U$) for $\Rey=3600$, $\Rey=4300$ and $\Rey=4700$ in the center
of region I at the point $(x,y)=(0.17\,d,0.31\,d)$. This point has been selected, because
the flow instability can be measured sensitively in this region. Shown is a period of
12\,s, equivalent to approximately one full revolution of the sphere around the torus. At
$\Rey=3600$ (figure \ref{fig:onefullrevolution}a) the flow is fully
developed, except for the immediate vicinity of the sphere where the streamwise velocity component varies
strongly. In some distance from the sphere the measured velocity $w$ is constant
up to small fluctuations which are mainly due to the noise of the PIV data.
Excluding the range disturbed by the sphere, the RMS value of the streamwise velocity is
$0.78\%$ of the mean value.

At $\Rey=4300$ (figure \ref{fig:onefullrevolution}b) a very distinct sinusoidal
modulation of the streamwise velocity is observed which differs markedly
from the signature of the wake flow. With the center of the sphere at
$\varphi=0$ the oscillation amplitude is nearly constant in a certain region around
$\varphi=\pi$. To analyze the oscillatory flow in the bulk we consider only a
time interval of 2.25\,s corresponding to 0.5\,m or $\simeq \pi/2$ which leaves a sufficient safety margin
from the region disturbed by the sphere. This range is marked as
\textit{analyzed} in figure \ref{fig:onefullrevolution}b and it is typically located on the side of the torus opposite to
the sphere. At $\Rey=4700$ a dominant frequency can still be recognized, but the
amplitude is no longer constant.

\begin{figure}
  \centerline{\includegraphics[clip,width=1\textwidth,angle=0]{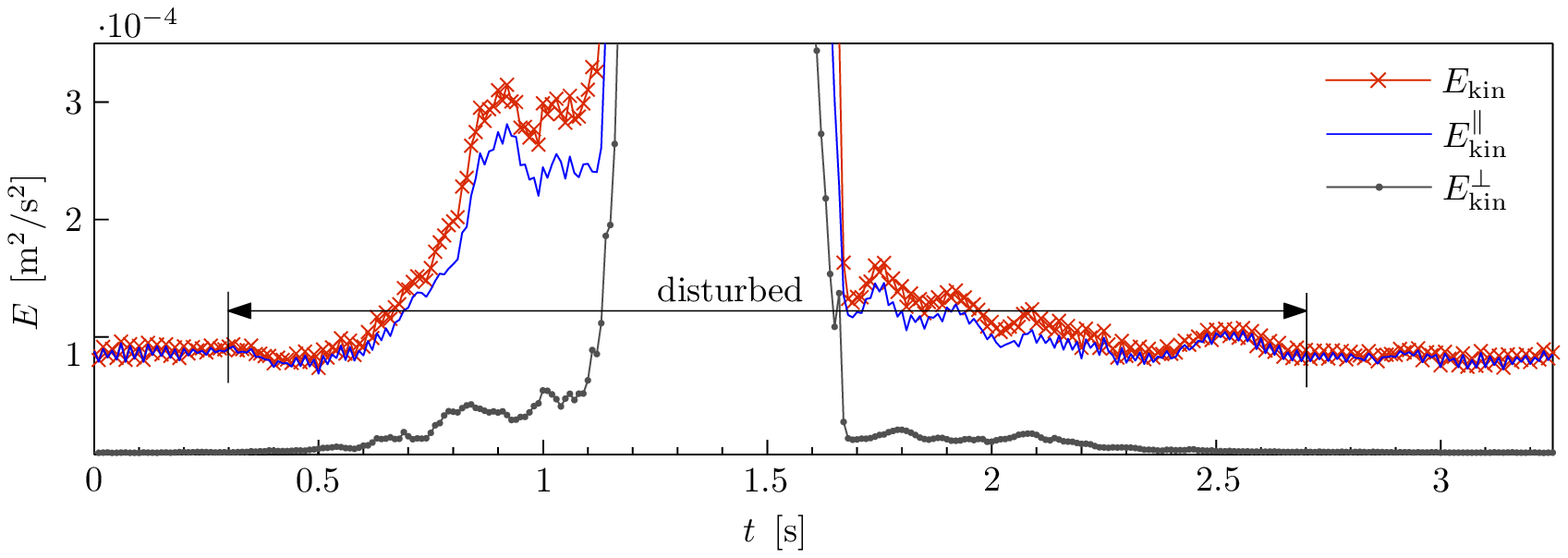}}
  \caption{\label{fig:Re=3000_Ekin_Transit}Transit of the sphere through the measurement
  plane for $\Rey=3600$. The spatially averaged (over $x$ and $y$)
  kinetic energy of the velocity fluctuations is displayed the direct vicinity of the sphere.
  Shown are the streamwise $(E_{\rm kin}^\|)$, the cross-stream $(E_{\rm
  kin}^\perp)$ and the total fluctuation energies $E_{\rm kin}= E_{\rm kin}^\| + E_{\rm kin}^\perp$.}
\end{figure}

To confirm that the perturbations induced by the rolling sphere are spatially restricted
we consider the kinetic energy. Decomposing the velocity field into a temporal
mean and a fluctuation part
$\vec u = \bar{\vec u} + \vec u'$ we define the streamwise and cross-stream fluctuation
energies $E_\| = \braket{w'^2}$ and $E_\perp = \braket{u'^2 + v'^2}$, respectively, where
the brackets indicate averaging over the $(x,y)$-plane. Figure
\ref{fig:Re=3000_Ekin_Transit} shows that the range of perturbed flow is confined to the
vicinity of the sphere as indicated by the arrows, while further away from the sphere the
kinetic energy of the fluctuations is nearly constant and very small. For $\Rey=3600$ the
flow is perturbed for $\approx 2.4$\,s during passage of the sphere, corresponding to an
azimuthal angle of $0.46\,\pi$ slightly asymmetric about the location of the sphere.
It is concluded that there exists a range of fully developed flow which is nearly
unperturbed by the details of the flow around the rolling sphere.

As the Reynolds number increases the unperturbed range decreases. Based on an
extrapolation of the measured length of the zone disturbed by the sphere at low and moderate Reynolds
numbers the length of the zone of the nearly fully-developed bulk flow for $\Rey=6000$ is at least $15\%$
of one full revolution of the torus. This range is predicted at $\varphi
\in [0.85\pi,1.15\pi]$ relative to the sphere at $\varphi=0$.
Even though the lengths of the wakes upstream and downstream from the
sphere differ, we selected a symmetric region for simplicity, leaving a sufficient safety
margin to the perturbed regions. All data analyzed were taken from these unperturbed
angular sections.

\subsubsection{Plunger with sphere}\label{subsec:plunger}

\begin{figure}
  \centerline{\includegraphics[clip,width=0.44\textwidth,angle=0]{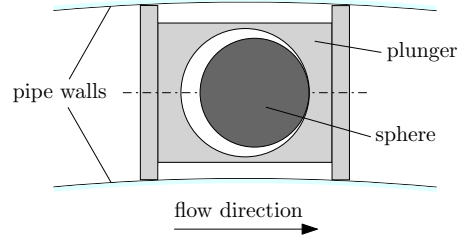}}
  \caption{\label{fig:plunger}Sketch of the first type of plunger within the torus.
  A small sphere of $d_{s}=18$mm placed inside the plunger is used as object of magnetic
  attraction to move it. Only the streamwise moving walls of the plunger, instead of the
  surface of the rolling sphere, interact with the flow in the torus.}
\end{figure}

To assess the effect of the rolling of the sphere on the end-wall-induced perturbations
and possibly on the bulk flow the driving was modified. Instead of a sole sphere a
plunger was used made from polyoxymethylene (POM) (Figure \ref{fig:plunger}). The plunger was machined in form of a slightly modified
cylindrical block with a diameter of 30.1\,mm by which the total area of the gap was
reduced to 9.49\,mm$^{2}$ or $1.3$\% of the tube's cross section.

A bore of 20\,mm diameter was drilled into the plunger, providing space for a steel
sphere with 18\,mm diameter. This steel sphere was also controlled by the magnet from
outside, but its rolling motion would not influence the flow in the torus. As the plunger
was cylindrical, its shape did not perfectly match the slightly curved torus. Due to the
lubrication of the fluid between the two solid surfaces of tube and plunger, the plunger
would slide through the tube very well without excessive stresses or seizures at the
wall. Only very mild wear, mostly originating from the tracer particles in the fluid, was
noticed.

We found that the design type of the actuator had very little influence on the length of
the perturbation flow near the moving end wall and no influence on the bulk flow. For
that reason only a rolling sphere was used for all measurements.

\subsubsection{Leak flow past the sphere}\label{subsubsec:leakflow}

The azimuthal velocity $U_s$ of the center of the sphere is precisely known as it is
controlled by the angular velocity $\Omega$ of the rotating boom. Comparing the
associated flow rate $U_s \pi d^2/4$ with the measured flow rate allows to determine the
leak flow past the sphere. The actual mean velocity in the torus $U$ was determined in
two ways. First, the pressure drop $\Delta p$ over the angle $\Delta\varphi = \pi/2$ was
measured. The result was compared to pressure-drop data obtained from published
correlations of \cite{White1929,Hasson1955,Mishra1979} for the laminar regime, which all deviate less than 1.5\%. Up to $\Rey\lesssim2500$ we find excellent qualitative agreement with these correlations but a practically constant deviation which can be attributed to a slightly slower velocity of the fluid $U$ than the azimuthal velocity $U_s$ of the sphere by $\approx3$\%. In addition, $U$ was obtained from SPIV-measurements. The resulting leak flux amounted to $3$\% of the actual mean volume flux, consistent with the flow rate from the pressure drop. Therefore, we can safely determine the mean velocity as $U = 0.97\,U_s$. It must be noted, however, that a slightly non-linear behavior of the leak flow for increasing Reynolds number could not be ruled out.

\subsection{Velocity-field measurements}\label{subsec:flowfield}

\subsubsection{Steady basic flow}\label{subsubsec:basicflow}

\begin{figure}
  \centerline{\includegraphics[clip,width=0.8\textwidth,angle=0]{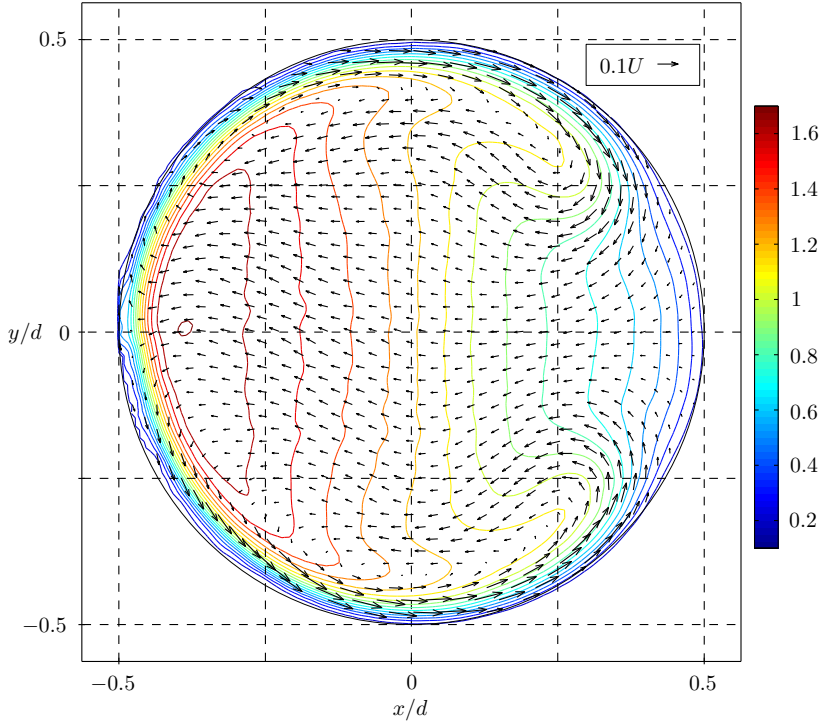}}
  \caption{\label{fig:mean_Re3000_rot-norm_colorcontour} Entire cross-sectional flow field
  of the tube in the steady laminar regime (at $\Rey=3600$). Streamwise (color contours)
  and in-plane velocity (vectors), both normalized with the mean streamwise velocity $U$.
  To eliminate measurement noise, the flow field is an average of 400 PIV-images.}
\end{figure}

In the following the flow field in the bulk is considered which is practically unaffected by
the end-wall perturbations. For small Reynolds numbers the flow field is
unique and steady. For $\Rey=3600$ the steady basic flow has developed its characteristic
structure which is caused by the pipe curvature. The velocity field measured by SPIV in
the entire cross-section is shown in figure
\ref{fig:mean_Re3000_rot-norm_colorcontour}. As the fluid in the core region is centrifugally driven toward
the outer wall, the maximum streamwise velocity is located in the plane of symmetry and
near the outer wall. Therefore, the gradient of the streamwise velocity component $u_\|$
is very high near the outer wall. As a result of the secondary, radial
outward motion in a wide range about the symmetry plane and due to continuity, fluid with
high streamwise momentum is transported radially inward along very thin cross-stream wall
layers. This is visible from the cross-stream velocity component $u_\perp$ and the
elongated contour shapes of the streamwise velocity. As the wall jets evolve radially
inward they widen. At approximately $\alpha=\pm \pi/4$ the jets turn and merge with the
broad radially outward stream, without separating from the wall. The two symmetrically
located turning regions of the secondary flow, denoted region I, represent the vortex
cores of the two Dean vortices.

The stationary waviness of the radial outward cross-stream flow near the equatorial line
in figure \ref{fig:mean_Re3000_rot-norm_colorcontour} indicates a weak symmetry breaking.
It could not be determined conclusively whether the small loss of symmetry is due to the flow physics or due
to systematic measurement errors. Possible sources of error are the symmetry breaking by
the rolling of the sphere and an imperfect calibration due to the custom made calibration
target (see sec.\ \ref{subsec:measurements}). However, since the cross-stream velocity is
about one order of magnitude smaller than the streamwise velocity it is evident that the
deviation from perfect symmetry is diminutive.

\begin{figure}
  \centerline{\includegraphics[clip,width=0.54\textwidth,angle=0]{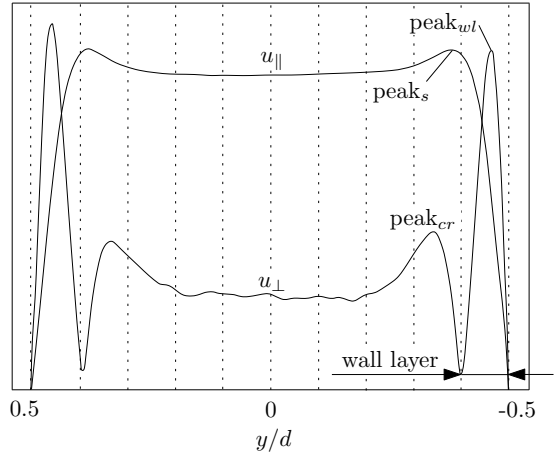}}
  \caption{\label{fig:profile_Re2000_vergleich}Vertical $(x=0)$ profiles of the streamwise
  ($u_\|$) and in-plane ($u_\perp$) velocity magnitudes for $\Rey=2400$. Velocities not to scale.}
\end{figure}

Figure \ref{fig:profile_Re2000_vergleich} shows profiles along the vertical $x=0$ of the
streamwise velocity $u_\|=w(y)$ and the absolute value of the cross-stream velocity
$u_\perp=\sqrt{u^2(y)+v^2(y)}$ ($\approx |u(y)|$ for $x=0$) for $\Rey=2400$. The wall
layers are clearly visible with global maxima of the absolute value of the cross-stream
velocity at $y=\pm 0.46\,d$ in a distance of $0.04\,d$ from each wall. The two other
maxima of the absolute value of the cross-stream velocity at $y=\pm 0.34\,d$ mark the
edge of the cross-stream return flow in the interior. Between these extrema of $u_\perp$ the
streamwise velocity $u_\|$ reaches its extrema at $y=\pm0.38\,d$ before dropping sharply
towards the wall. The boundary layers become thinner with increasing Reynolds number
and/or curvature of the pipe \citep[see also][]{Webster1997}.

\subsubsection{Mean-velocity profiles for increasing Reynolds
number}\label{subsubsec:meanvelocityincre}

\begin{figure}
  \centerline{\includegraphics[clip,width=0.8\textwidth,angle=0]{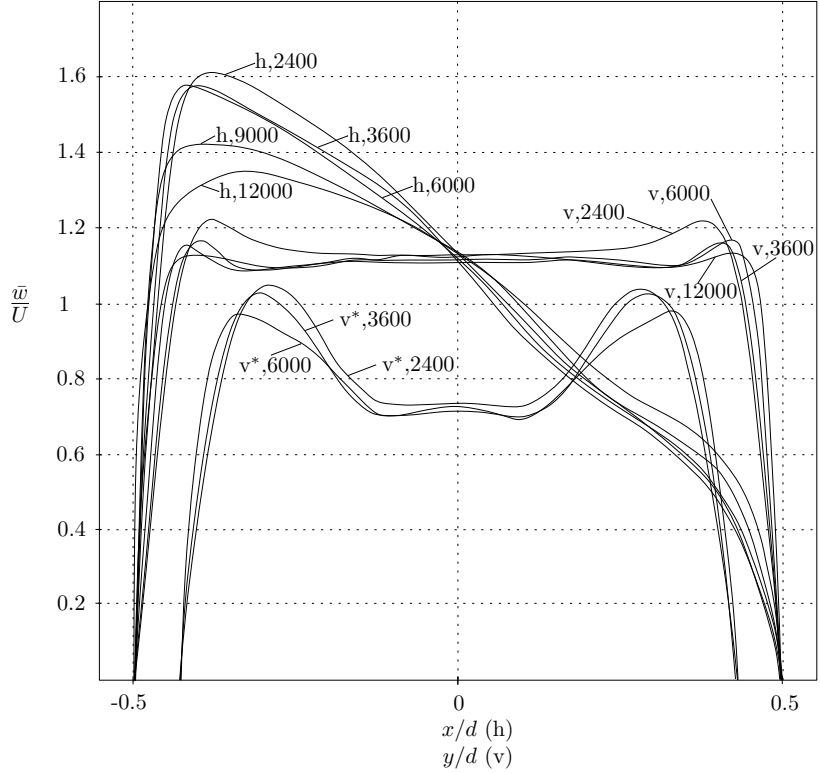}}
  \caption{\label{fig:profiles_hor-vert}Horizontal (h) and vertical (v) profiles of the
  mean streamwise velocity at $y=0$ and $x=0$, respectively, for different Reynolds
  numbers. v$^\ast$ denotes vertical profiles along $x=0.25\,d$. }
\end{figure}

For higher Reynolds numbers the basic flow becomes unstable and time-dependent. To
characterize the flow in a wide range of Reynolds numbers we consider the
time-averaged streamwise velocity $\bar
u_\|=\bar w/U$ in the two orthogonal planes $x=0$ and $y=0$. Figure
\ref{fig:profiles_hor-vert} shows the profiles $\bar u_\|(x,0)$ (denoted by h for
horizontal) and $\bar u_\|(0,y)$ (denoted by v for vertical) for various Reynolds
numbers. Characteristic for curved-pipe flow are the flat vertical profiles and the
nearly linear variation of the horizontal profiles. As the Reynolds number increases the
boundary layers become thinner, associated with increasing strain rates at the walls.
Moreover, the slope of the horizontal velocity profiles becomes smaller, associated with
a reduction of the velocity extrema which is most significant for $\Rey=O(10^4)$. It is
remarkable that the horizontal profiles of the mean velocity do not change very much
between $\Rey=3600$ and $\Rey=6000$, even though the flow for $\Rey=6000$ is strongly
time-dependent. In addition to the profiles of the mean streamwise velocity along $y=0$
and $x=0$, $\bar u_\|(0.25\,d,y)$ (denoted by v$^*$) is shown. As can be seen, the mean
velocity profiles along $x=0.25\,d$ (v$^\ast$) do experience a change between $\Rey=3600$
and $\Rey=6000$. This change is related to the the onset of oscillatory flow whose
amplitude is largest near the maxima of $\bar u_\|$ along $x=0.25\,d$.

\subsection{Primary instability: The onset of time-dependence}\label{subsec:transition1}

\begin{figure}
  \centerline{\includegraphics[clip,width=0.9\textwidth,angle=0]{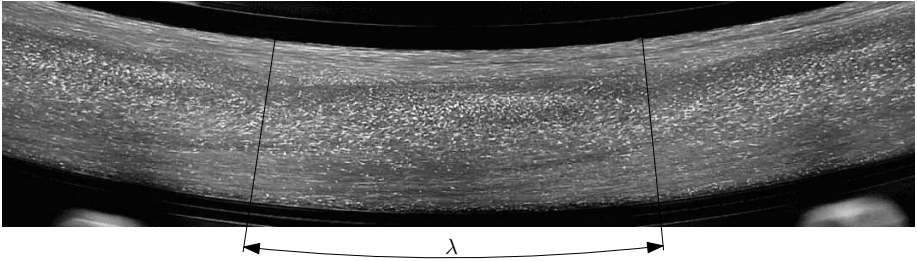}}
  \caption{\label{Foto_twave3}Still picture from a movie (see
  \textit{mov-Re4350.1turn.avi}) of a short section of the torus from above
  at $\Rey=4350$. The camera was following the flow with the velocity of the sphere $U_s$.}
\end{figure}

When the Reynolds number is increased the steady basic flow becomes unstable and
time-dependence sets in at $\Rey_c=4075$. The supercritical flow arises as a
wave which travels downstream with a phase velocity which is slightly larger than the
mean flow. With increasing Reynolds number the amplitude grows continuously from zero at
the threshold. Figure \ref{Foto_twave3} shows a short section ($\Delta
\varphi\cong0.13\pi$) of the torus at $\Rey=4350$. The wavelength $\lambda$ is indicated
by arrows. The wave can be recognized much better in the short movie \textit{mov-Re4350.1turn.avi} (supporting material, online) than in the image.

\begin{figure}
  \centerline{\includegraphics[clip,width=0.54\textwidth,angle=0]{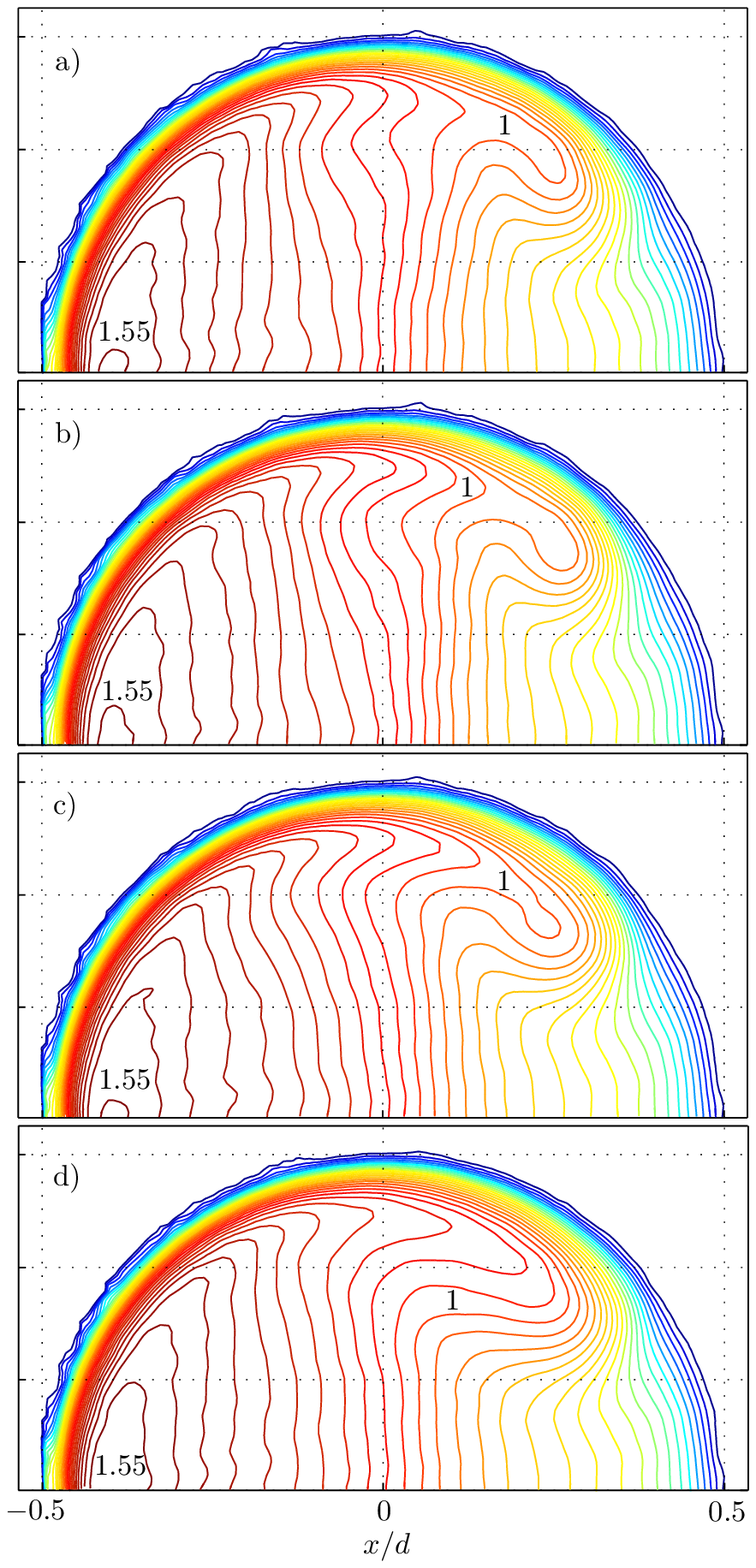}}
  \caption{\label{fig:Re4300_totvel_4subpic} Contours of the (instantaneous) streamwise
  velocity $w$ over one wavelength $\lambda$ at $\Rey=4300$ in increments of $0.05\,U$. a)
  $\lambda=0$, b) $\lambda=1/4$, c) $\lambda=1/2$ and d) $\lambda=3/4$. The level of $1.55\,U$
  and $1\,U$ is indicated for reference.}
\end{figure}

\begin{figure}
  \centerline{\includegraphics[clip,width=0.9\textwidth,angle=0]{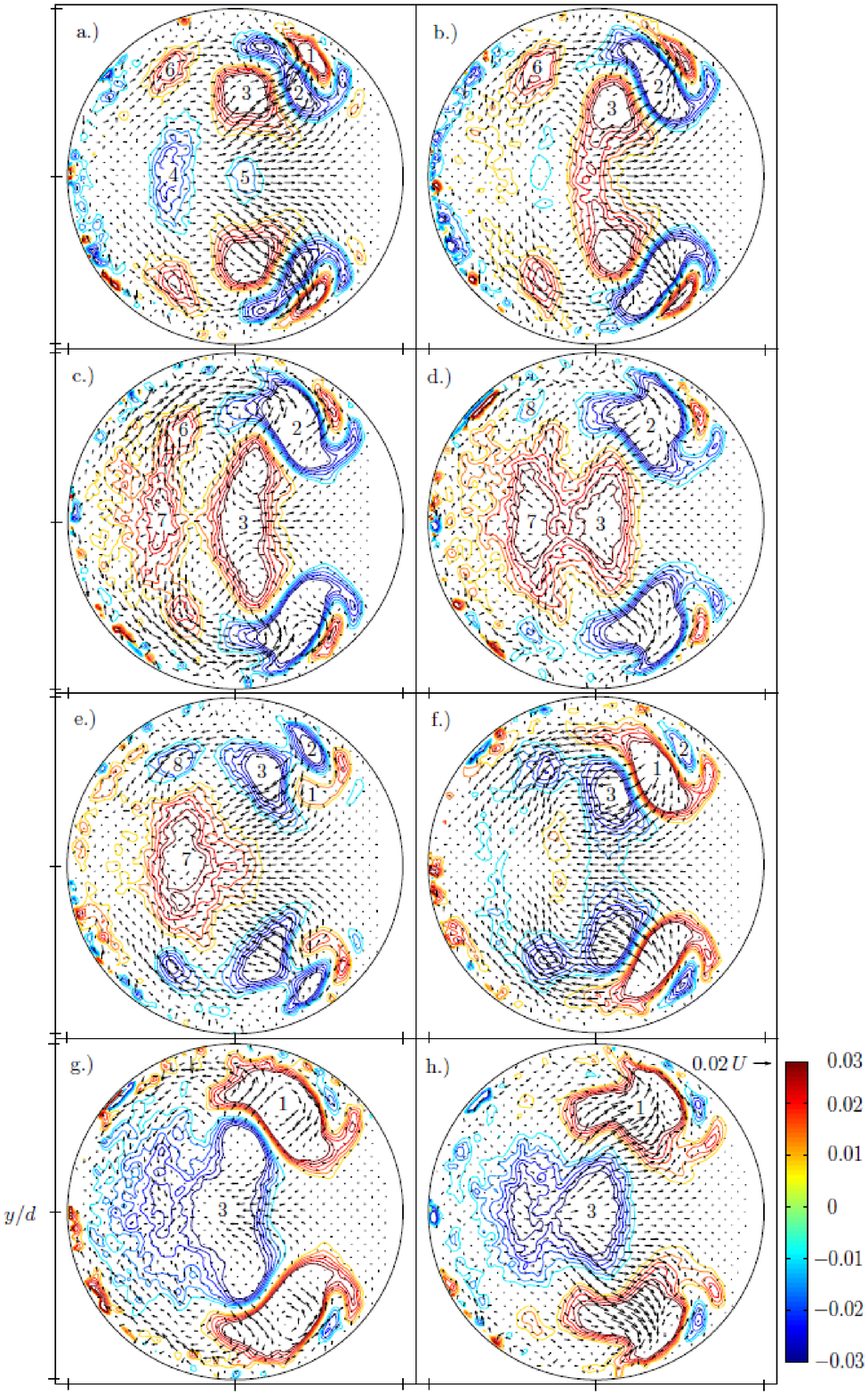}}
  \caption{\label{fig:Re4300_phaverfluctvel_8subfigs_all}Fluctuating velocities over one
  wavelength $\lambda$ at $\Rey=4300$. Vector field of the cross-stream velocity
  fluctuations and contours of the streamwise velocity fluctuations, both normalized with the bulk
  velocity $U$. For the sake of clearer representation $w'$ is displayed only for $\pm0.008$
  to $\pm0.03$ in increments of 0.004. The maximum levels would be $\pm0.065$.}
\end{figure}

From visual observations the wavelength $\lambda$ of the finite-amplitude wave in terms
of the arclength along the center of the pipe can be estimated as $\lambda =
\Delta\varphi D/2 \approx (0.06$--$0.08)\pi D/2 \approx (2$--$2.5)\,d$ for
$\Rey\leqq4350$. In the range of $4075\le \Rey \le 4350$ the wave is almost stationary in a frame of reference
moving with the sphere. The wave celerity $c=(1.1$--$1.13)\,U$ is about 10\% higher than
the mean velocity. For $\Rey\gtrapprox4350$ the strict periodicity of the wave is lost
and the peak-to-peak distance of the disturbed wave is varying between $2$--$4\,d$.
Moreover, for Reynolds numbers $\Rey\gtrapprox 4580$ the wavy motion is interrupted by
short irregular bursts. For even higher Reynolds numbers, $\Rey \gtrapprox 5050$, no
dominant period can be detected by visual observation.

At $\Rey=4300$ the finite-amplitude wave is already well established. Figure
\ref{fig:Re4300_totvel_4subpic} shows the streamwise velocity $w$ in the $(x,y)$-plane at
four instants during one period. Each image is obtained from averaging eight successive
periods. The oscillation amplitude of the cross-stream velocity is very small (not
shown). The fully time-resolved oscillatory flow is shown in
\textit{Re4300.timeresvelfield.gif} (supporting material, online). The oscillations are most pronounced in the
vicinity of the contour level $1\,U$, in particular in region I (see figure
\ref{fig:Referenzsystem_figures}). During the oscillation the finger of high streamwise
momentum in region I is oscillating back and forth in meridional direction (parallel to
the wall). When the fingers elongate towards point $A$ at the inner wall, the radially
stratified streamwise momentum in the center of the pipe moves towards point $C$ on the
outer wall, the shift however being smaller, and vice versa.

\begin{figure}
\begin{center}
  \subfigure[View from the inside of the
  bend]{\includegraphics[clip,width=0.71\textwidth]{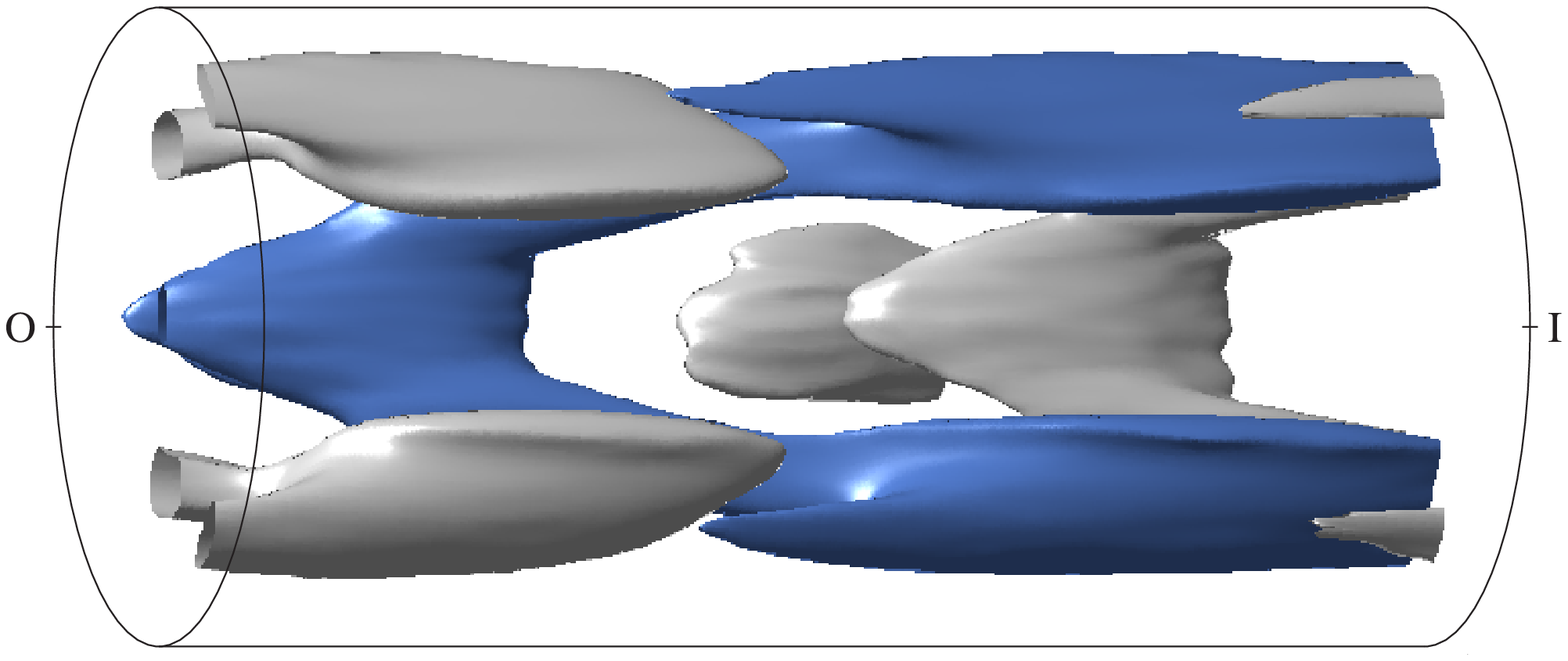}}\qquad
  \subfigure[View from the outside of the
  bend]{\includegraphics[clip,width=0.71\textwidth]{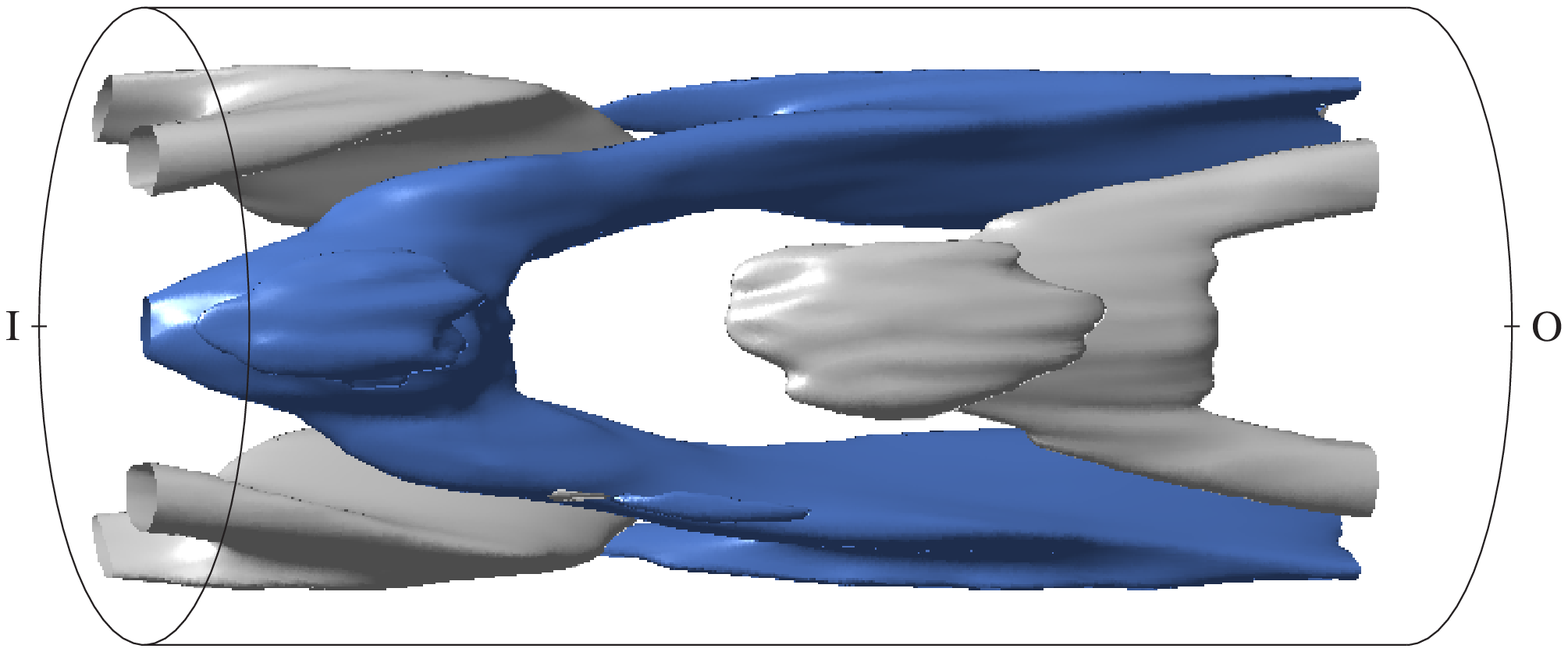}}\\
\end{center}
  \caption{\label{fig:Re4300_isosurface}Isosurfaces of streamwise velocity oscillations
  (streaks) at $\pm0.028\,U$ for $\Rey=4300$ shown over one wavelength $\lambda$.
  Two views from opposing sides (I denotes inner wall, O outer wall) are shown.
  The flow is from left to right. Blue indicates a negative, light gray indicates a
  positive streamwise velocity perturbation. The images have been averaged of
  eight periods. For reasons of representation, the velocity field is mapped to a
  straight pipe.}
\end{figure}

Figure \ref{fig:Re4300_phaverfluctvel_8subfigs_all} shows the vector field of the
cross-stream velocity fluctuations and selected contour lines of the streamwise velocity
fluctuations at eight instants during one period (see also
\textit{Re4300.fluctvelocity.gif} in the supporting material, online). Each image is
obtained from averaging 8 periods of oscillation. Despite of the slight asymmetry of the
basic flow the oscillations (deviations from the temporal mean) are symmetric with
respect to the horizontal plane. The magnitudes of the velocity oscillations in the
streamwise and the cross-stream directions are of the order of $\pm0.06\,U$ and
$\pm0.02\,U$, respectively. The cross-stream velocity field in (a) consists of a radial
outward flow from the center (labeled 5) of the pipe in the right half of the cross
section. The radially outward flow at $\alpha\approx\pi/4$ towards and across region I is
particularly strong (point 2). Near point 6 at the edge of the cross-stream wall jet at
$\alpha\approx 3\pi/4$ the cross-stream velocity is oriented tangentially in negative
$\alpha$-direction. The cross-stream oscillation amplitude is very small in the regions
of low and high streamwise basic-state velocity near the equatorial points A and C,
respectively. Half a period later (e) the cross-stream-flow field is reversed. These
cross-stream velocity perturbations act on the underlying mean streamwise velocity field
(basic flow). Owing to the cross-stream gradients of the mean streamwise flow the
cross-stream perturbation flow creates the streamwise perturbations (labeled by numbers
in figure \ref{fig:Re4300_phaverfluctvel_8subfigs_all}) which can be considered as
streaks. The periodic cross-stream velocity acts, in particular, on the fingers of the
streamwise mean flow. Since the streamwise mean velocity exhibits a minimum and a maximum
along the ray $\alpha=\pi/4$ (see figure \ref{fig:Re4300_totvel_4subpic}) the radially
outward cross-stream perturbation flow along this direction creates three streaks (1,2,3)
of alternating sign. During the temporal (or spatial) evolution, the central streak (2)
grows while the other streaks move radially inward (3) and outward (1). After streak (2)
has grown to a considerable size (figure \ref{fig:Re4300_phaverfluctvel_8subfigs_all}d)
it splits and the original streak structure (figure
\ref{fig:Re4300_phaverfluctvel_8subfigs_all}a) is recovered in figure
\ref{fig:Re4300_phaverfluctvel_8subfigs_all}e, albeit with a different sign. The pair of
radially inward moving streaks merge at the equatorial plane (figure
\ref{fig:Re4300_phaverfluctvel_8subfigs_all}b--c) and subsequently merge
with the two merged streaks (6) which have also moved towards the equatorial plane.

The time- and space-resolved structure of the streamwise velocity fluctuation is shown in
figure \ref{fig:Re4300_isosurface} for $\Rey=4300$. Isosurfaces at $\pm0.028\,U$ are
shown from the inside (a) and from the outside of the bend (b). The streaks, i.e.\ the
regions where the streamwise velocity is lower or higher than the steady laminar value,
are shown in blue and light gray, respectively. The mirror symmetry with respect to the
equatorial plane is obvious.

\begin{figure}
  \centerline{\includegraphics[clip,width=0.5\textwidth,angle=0]{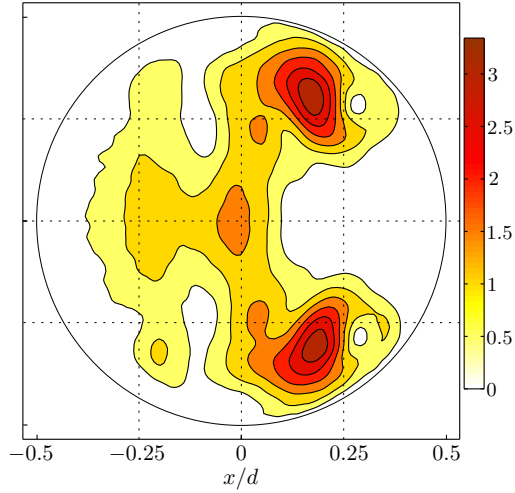}}
  \caption{\label{fig:Re4300_Amplitudes}Spatial distribution of the amplitude of the
  fundamental Fourier mode $A_1(x,y)$ in the streamwise velocity-fluctuation
  field for $\Rey=4300$ measured by PIV. The amplitude is given in arbitrary units.}
\end{figure}

The spatial distribution of the fundamental Fourier mode $A_1(x,y)$ with frequency $f_1$
is shown in figure \ref{fig:Re4300_Amplitudes} for the streamwise velocity fluctuations
$u_\|$. The fundamental mode is dominating at $\Rey=4300$ and the corresponding
streamwise velocity fluctuation is sharply peaked at $(x,y)=(0.17\,d,\pm0.31\,d)$. Near
the center another region of streamwise velocity perturbation is found which is quite
wide, but with smaller amplitude. The streamwise velocity fluctuation is practically
absent in the low-speed core region and in the cross-stream wall layers. This result
underlines the importance of region I for the instability, consistent with the
observation of \cite{Webster1997} who, likewise, found a maximum oscillation amplitude in
this region of the flow.

\begin{figure}
\begin{center}
  \subfigure[$\Rey=4350$]{\includegraphics[clip,width=0.47\textwidth]{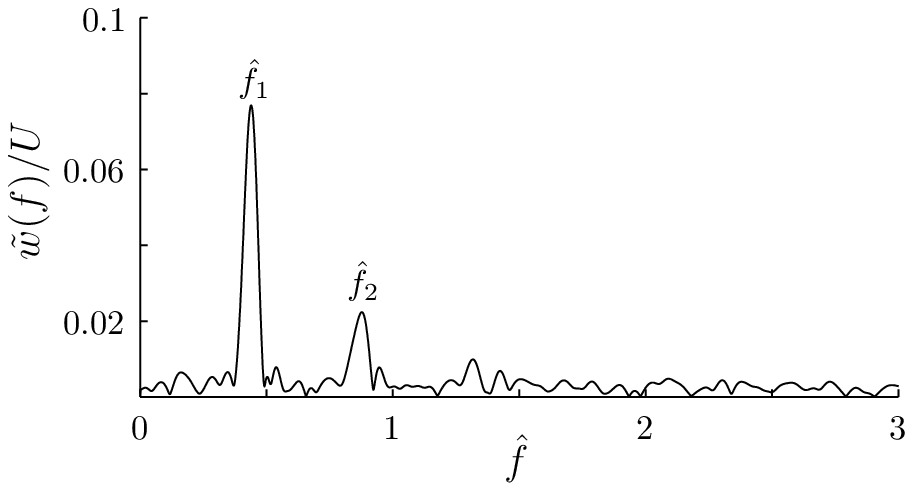}}\qquad
  \subfigure[$\Rey=4400$]{\includegraphics[clip,width=0.47\textwidth]{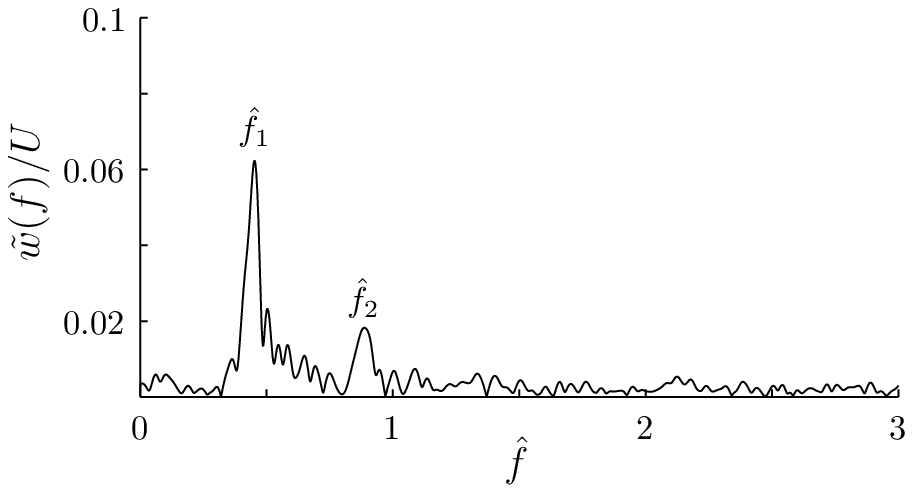}}\\
  \subfigure[$\Rey=4700$]{\includegraphics[clip,width=0.47\textwidth]{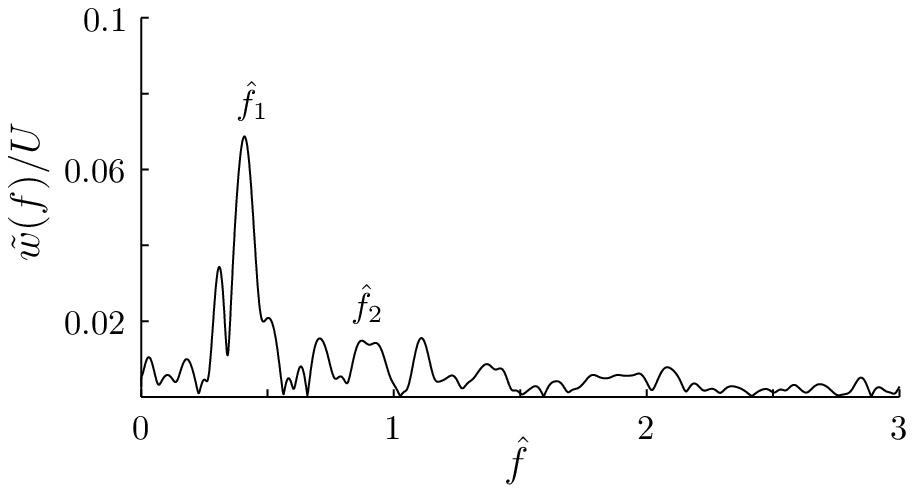}}\qquad
  \subfigure[$\Rey=5000$]{\includegraphics[clip,width=0.47\textwidth]{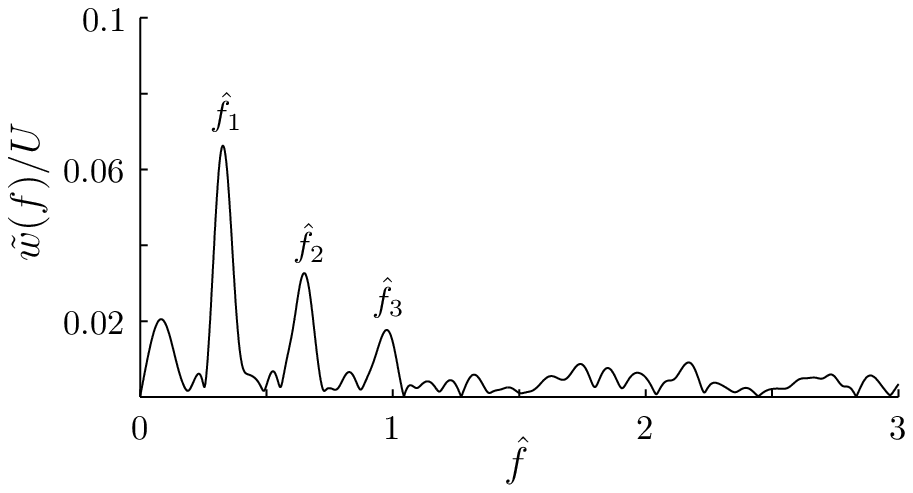}}
\end{center}
  \caption{\label{fig:Spectra}Spectra of $w$ measured by LDV at $(x,y)=(0.17\,d,\pm0.31\,d)$
  obtained from $\varphi=[0.85\pi,1.15\pi]$ during a single revolution of the sphere.
  Shown are the Fourier amplitudes of the streamwise fluctuations
  normalized with the mean velocity in the torus, i.e.\ $\tilde w(f)/U$. }
\end{figure}

The absolute maximum of the velocity perturbation in region I provides the best
signal-to-noise ratio for measuring the streamwise velocity. Since the location of the
maximum was nearly independent of the Reynolds number, the position
$(x,y)=(0.17\,d,0.31\,d)$ was selected for LDV measurements. The measured streamwise
velocity perturbation $w$ was Fourier analyzed for different Reynolds numbers. Figure
\ref{fig:Spectra} shows spectra 
at four different Reynolds numbers obtained from the range
$\varphi=[0.85\pi,1.15\pi]$ ($\varphi=0$ corresponds to the location of the sphere)
during a single revolution of the sphere.
The dimensionless frequency (Strouhal number) is obtained as $\hat{f} = fd/U$, where $f$ is the frequency in Hz.
At $\Rey=4350$ (a) a fundamental frequency $\hat{f_1}$ and its weak second harmonic frequency $\hat{f_2}=2\hat{f_1}$ is clearly resolved.
At $\Rey=4400$ (b) and $\Rey=4700$ (c) the same fundamental frequency $\hat{f_1}$ is still dominant, but its
amplitude is less than for $\Rey=4350$. Furthermore, additional frequencies in the vicinity of $\hat{f_1}$ can be detected. At $\Rey=5000$ (d) no additional frequencies apart from $\hat{f_1}$ and its higher harmonics can be resolved.

The spectrum for $\Rey=4350$ (figure \ref{fig:Spectra}a) is reproducible for each
revolution of the sphere and it is characteristic for the range $4075\leq\Rey\leq4350$.
The amplitudes $\tilde w_{1,2}$ differ slightly from revolution to revolution. The
dominant frequency $\hat{f_1}$, however, remained constant. For cases (b), (c) and (d) the
measured spectra varied distinctly from revolution to revolution (not shown).

\begin{figure}
  \centerline{\includegraphics[clip,width=0.96\textwidth,angle=0]{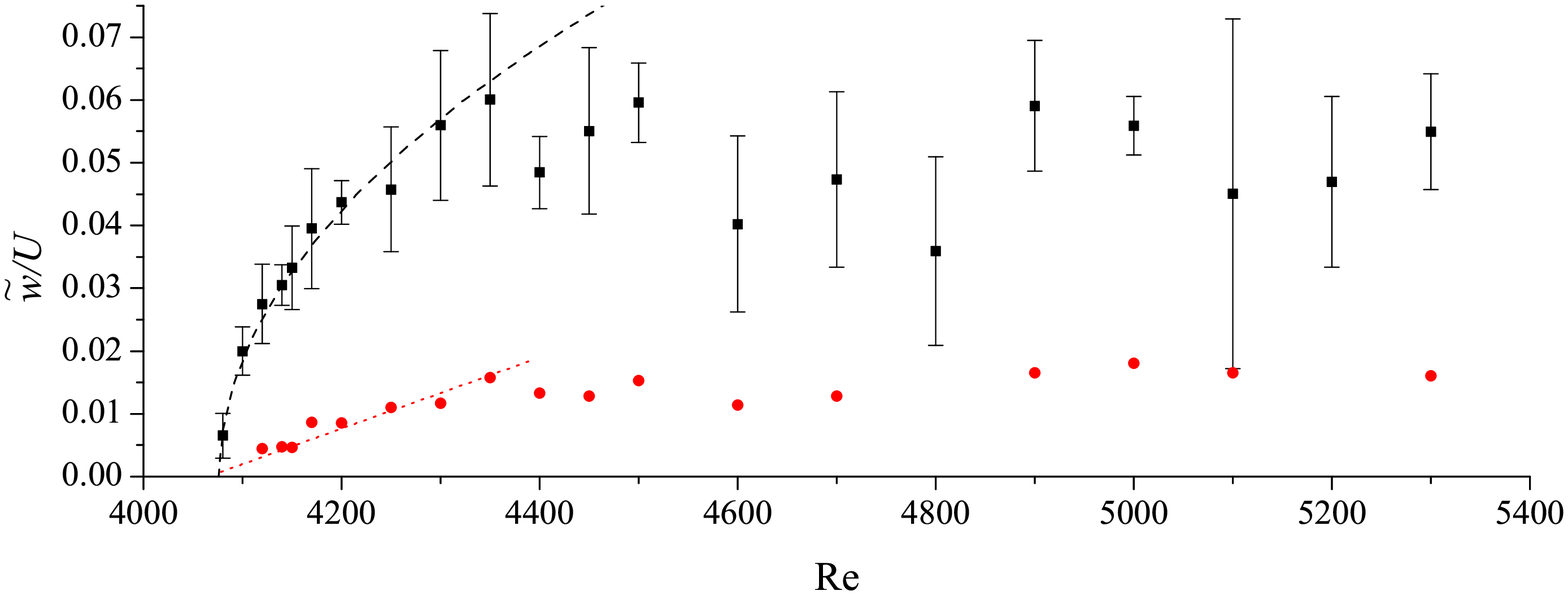}}
  \caption{\label{fig:A_vs_RE}Normalized  amplitude $\tilde w_1/U$ of the dominant frequency $\hat{f_1}$ (squares) and
  its harmonic $\tilde w_2/U$ (dots) for increasing Reynolds number. Error bars represent the maximum deviation of
  the measured amplitude for single revolutions. The dashed and dotted lines represent square-root and linear approximations, respectively.}
\end{figure}

The first occurrence of $\hat{f_1}$ within the velocity spectrum indicates the first critical Reynolds number. The dependence of
the amplitude $\tilde w_1$ on the Reynolds number is indicative of
supercritical bifurcation. To reduce the deviations among the results
obtained from different revolutions, the angular range $[0.85\pi,1.15\pi]$ was Fourier
analyzed and the result averaged over five revolutions of the sphere for each Reynolds
number. The result is shown in figure \ref{fig:A_vs_RE}. The error bar
indicates the maximum deviation of an individual amplitude from the mean
value. Up to $\Rey=4350$ the measured data can be fitted to a square-root law (black dashed line)
\begin{equation}
  \tilde w_1(\Rey) = a_1\sqrt{\Rey-\Rey_{c1}}
\end{equation}
with $a_1=4.7\times10^{-3}$ and $\Rey_{c1}=4075$. The amplitude of the second harmonic
$\tilde w_2$ is also shown. This supercritical bifurcation for $\delta=0.049$ is in contrast to the subcritical
bifurcation for $\delta=0.1$ found by \cite{Piazza2011} numerically in the periodic torus.

\begin{figure}
  \centerline{\includegraphics[clip,width=0.96\textwidth,angle=0]{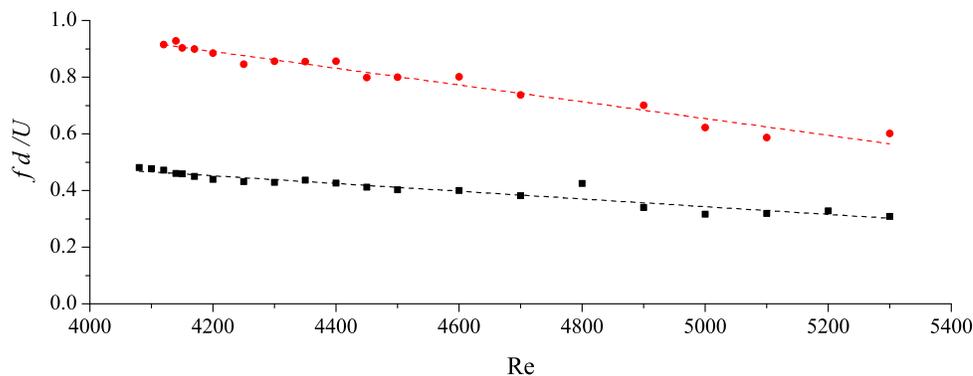}}
  \caption{\label{fig:F_vs_RE}Strouhal number $\hat{f_1}$ (black squares)
  and its harmonic $\hat{f_2}$ (red dots) as functions of the Reynolds number.  Linear fits are indicated as dashed lines.}
\end{figure}

Figure \ref{fig:F_vs_RE} displays $\hat{f_1}$ and $\hat{f_2}$ in the range
$4000\leq\Rey\leq5400$. The fundamental frequency (black squares) decreases slightly with the Reynolds number
and can be fitted to a linear curve (black dashed line in figure \ref{fig:F_vs_RE})
\begin{equation}\label{eq:linfitfrequency}
  \hat{f_1}=\frac{f_{1}d}{U}=1.05-1.42\times10^{-4}\Rey.
\end{equation}
Up to the measurement error this behavior is
also satisfied by the second harmonic frequency $\hat{f_2}$. Apart from $\hat{f_1}$ and
$\hat{f_2}$ no further dominant frequencies are found in the velocity spectra in the
range $\Rey_{c\,1}\leq\Rey\leq4350$. However, the fundamental frequency
$\hat{f_1}(\Rey)$ can be traced to much higher Reynolds numbers into the regime of more
complex flows in which periodicity in the bulk is lost.

\subsection{Beyond the first instability}\label{subsec:transition2}

The Hopf instability at $\Rey_{c\,1}=4075$ is breaking the continuous rotational
invariance with respect to $\varphi$ while preserving the mirror symmetry with respect to
the equatorial plane. The deviation of $\tilde w_1(\Rey)$ from the square root law in
figure \ref{fig:A_vs_RE} suggests that a secondary bifurcation occurs near
$\Rey_{c2}\approx 4400$. Around this Reynolds number the peak-to-peak distance of the
flow oscillations was visually observed to start varying between approximately $2d$ and
$4d$, indicating a loss of simple periodicity. The widening of the spectrum near $\hat{f_1}$ in
figure \ref{fig:Spectra}\,b and c indicates that a further bifurcation to a state with two
possibly incommensurate frequencies $\hat{f_1}$ and $\hat f'_1$ takes place. The second
frequency $\hat f'_1$ is conjectured to exhibit a value below $\hat
f_1$. Despite of careful investigation of the velocity spectra $\hat f'_1$ could not be pinpointed as every revolution of the sphere yielded different spectra.

\begin{figure}
\centerline{\includegraphics[clip,width=1\textwidth,angle=0]{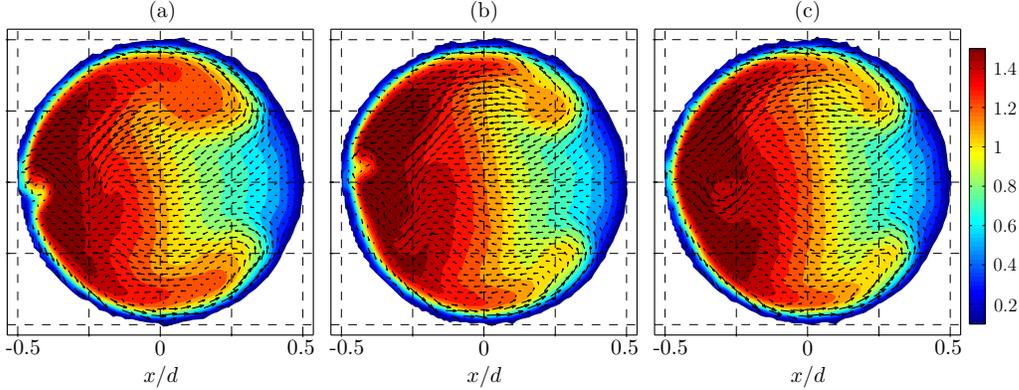}}
  \caption{\label{fig:Re4600_colorcontour_instantaneous3x}Cross sectional flow field for $\Rey=4600$ at different moments
  of time (same as in fig.\ \ref{Re4600_timetrace_4s_5points}): $t=0.60$ (a), $t=0.88$ (b) and $t=3.23$ (c). Note the small-scale vortex in the wall layer as well as the flow asymmetry in the interior.}
\end{figure}

Characteristic of the conjectured second instability is a breakdown of the mirror
symmetry with respect to the equatorial plane. This is confirmed in figure
\ref{fig:Re4600_colorcontour_instantaneous3x} which displays instantaneous flow fields at
$\Rey=4600$. The symmetric oscillations seem to 'break down' from time to time and,
associated with the break down, the symmetry with respect to the equatorial plane is lost
temporarily. The asymmetry is observed in the interior as well as in the wall layer. The
asymmetry in the interior arises in form of a non-zero velocity component $v(x,0)$ along
the $x$ axis in the outer half of the pipe $(x<0)$ and visible interior vortex structures for
$x<0$ which are asymmetric with respect to $y=0$.

\begin{figure}
  \centerline{\includegraphics[clip,width=0.98\textwidth,angle=0]{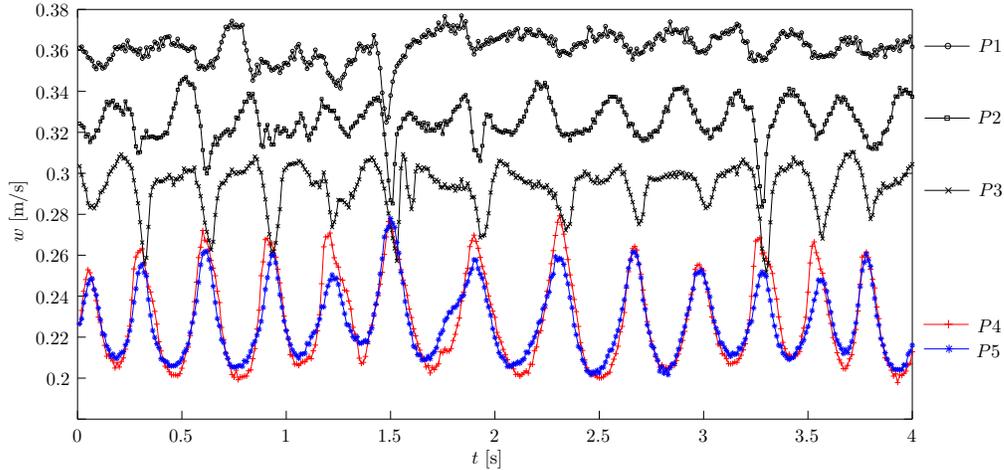}}
  \caption{\label{Re4600_timetrace_4s_5points}Time traces for $\Rey=4600$ of the streamwise
  velocity during four seconds (corresponding to $\approx 50$\% of one full revolution of the
  sphere or the range $\varphi\in[\pi/2,3\pi/2]$. Signals are shown for five locations $P1$ to $P5$ as
  explained in the text.}
\end{figure}

The equatorial symmetry is lost due to small-scale flow structures in the wall layers
near the equatorial plane and also by interior-flow structures. The small-scale vortices
arise spontaneously in the wall layer. One example is shown in figure
\ref{fig:Re4600_colorcontour_instantaneous3x}\,a. A low-velocity streak is deforming the isolines of the streamwise
flow in the wall layer just below the $x$-axis near point C. Such a streak is usually
caused by a pair of cross-stream vortices (only one of which can be identified in the
total flow in figure \ref{fig:Re4600_colorcontour_instantaneous3x}\,a). The structure is visible for a short period of
time of about 0.2\,s. Immediately after vanishing the same structure appears for the same
short period of time in mirror symmetrical location in the upper half of the torus
(figure \ref{fig:Re4600_colorcontour_instantaneous3x}\,b). In addition to these localized structures which seem to
arise erratically, we find asymmetric flows in the interior which seem to be related to the
underlying symmetric oscillations with frequency $f_1$. From time to time interior vortices
are created in the cross-stream flow in the outer half of the torus ($x<0$). An example
is shown in figure \ref{fig:Re4600_colorcontour_instantaneous3x}\,c. The asymmetric interior-flow structures do not seem
to be related to the small-scale vortices in the wall layer, but both can be observed for
$\Rey\gtrsim 4400$. Owing to the limited observation time we were not able to exactly
pinpoint a critical Reynolds number $\Rey_{c2}$ for the symmetry breaking.

\begin{figure}
\begin{center}
  \subfigure[$\Rey=6000$]{\includegraphics[clip,width=0.98\textwidth,angle=0]{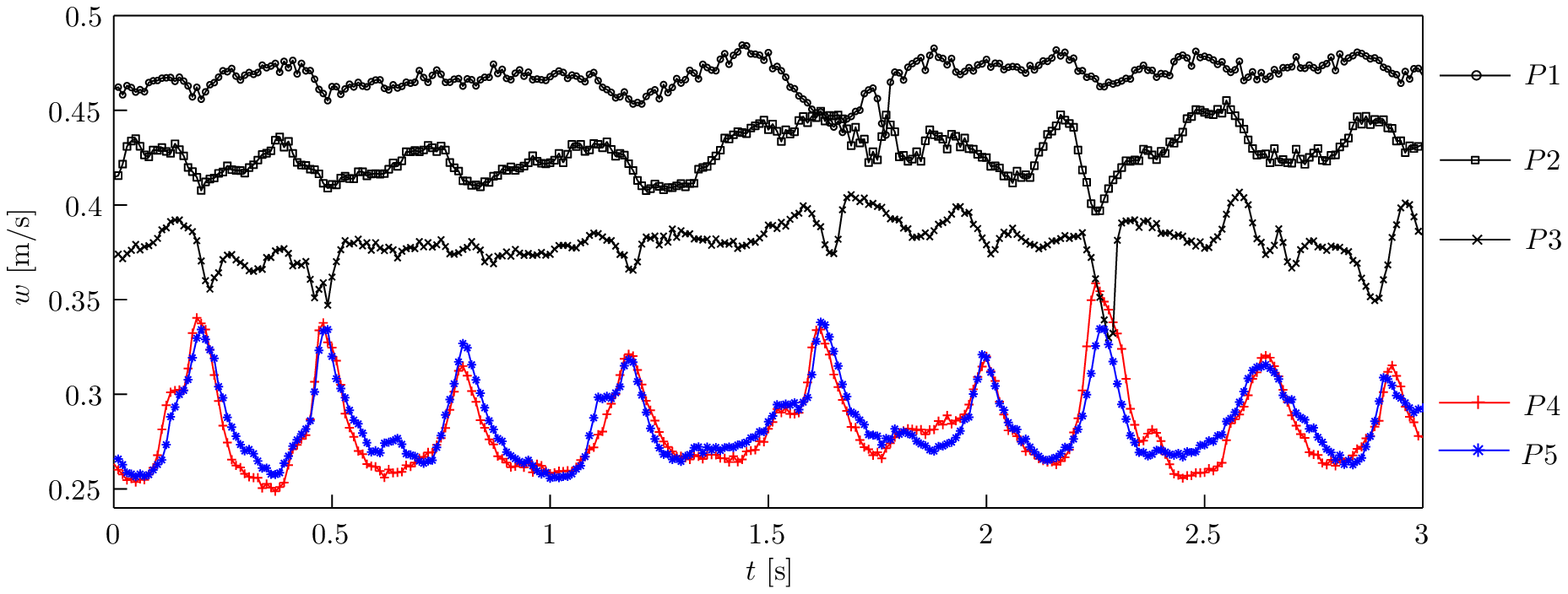}}\\
  \subfigure[$\Rey=9000$]{\includegraphics[clip,width=0.98\textwidth,angle=0]{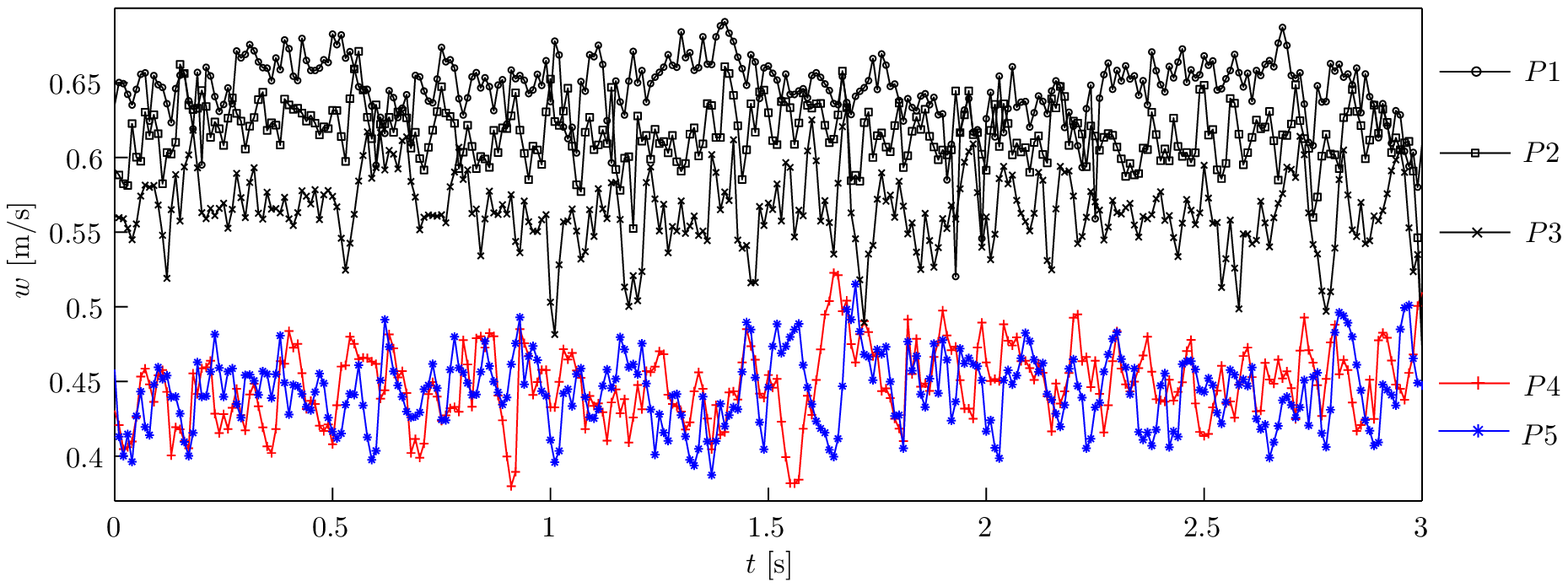}}
\end{center}
  \caption{\label{fig:timetraces_3s_5points}Time traces of the streamwise velocity during 3
  seconds (corresponding to $\sim50$\% of one full revolution for (a) and $\sim70$\% for
  (b)
  respectively) at $\Rey=6000$ (a) and $\Rey=9000$ (b) for 5 different points within the
  cross section. $P1-P5$ as in figure \ref{Re4600_timetrace_4s_5points}.}
\end{figure}

In order to further characterize the temporal and spatial behavior of the flow beyond the
second instability figure \ref{Re4600_timetrace_4s_5points} displays the streamwise
velocity $w(x,y,t)$ during a period of four seconds for $\Rey=4600$ corresponding to the
azimuthal range $\varphi\in[\pi/2,3\pi/2]$. Signals shown are obtained from three
locations in the mid-plane: $P1$: $(x_1,y_1)=(-0.38\,d,0)$, $P2$: $(x_2,y_2)
=(-0.25\,d,0)$, $P3$: $(x_3,y_3) =(-0.09\,d,0)$, and from two symmetrical locations in
region I: $P4$: $(x_4,y_4) = (0.17\,d,0.31\,d)$ and $P5$: $(x_5,y_5) =
(0.17\,d,-0.31\,d)$. For the periodic oscillatory flow for $\Rey_{c1}< \Rey < \Rey_{c2}$
all these signals are periodic as in figure \ref{fig:onefullrevolution}. Moreover, the
signal from the antisymmetric points $P4$ and $P5$ would be identical. For
$\Rey>\Rey_{c2}$ as shown in figure \ref{Re4600_timetrace_4s_5points} we observe the loss
of symmetry indicated by the differences between signals from $P4$ and $P5$. While the
flow is symmetric and periodic for a considerable time, it is interrupted by seemingly
random anti-symmetric and high-intensity bursts. Such bursts can be clearly identified in
the left half of the torus by the signals of $P1$, $P2$ and $P3$ at $t=1.5$ and of $P2$
and $P3$ at $t= 3.3$, leading to an asymmetric flow as indicated by the differences in
the signals from $P4$ and $P5$.

Figure \ref{fig:timetraces_3s_5points} shows the five signals at the same monitoring
points for $\Rey=6000$ (a) and for $\Rey=9000$ (b). For $\Rey=6000$ the time traces of
the streamwise velocity are similar to those at $\Rey=4600$. The symmetry of the signals
$P4$ and $P5$ is only slightly perturbed. However, the nonlinearity in the oscillations
is stronger which also shows up by higher Fourier components in the spectrum (not
shown). Furthermore, periodic oscillations are difficult to be recognized in the signals from
$P1$, $P2$ and $P3$.

At $\Rey=9000$ (figure \ref{fig:timetraces_3s_5points}b) the flow has changed
significantly as compared to $\Rey=6000$. The signal is dominated by
high-frequency and random oscillations in all monitoring points. The symmetry between
signals $P4$ and $P5$ is completely lost. The character of the fluctuations evidence
turbulent characteristics. It is noted that the overall amplitudes of the fluctuations
have decreased at $P4$ and $P5$ as compared to lower Reynolds numbers.

A comparison between figures \ref{fig:timetraces_3s_5points}a and b shows that the
high-frequency fluctuation in (b) are essentially absent in (a). Therefore, we consider
the flow at $\Rey=6000$ to be chaotic while the flow at $\Rey=9000$ is considered
turbulent. For a further classification of different flow regimes a more detailed
investigation of the turbulent fluctuations is required.

\subsection{Friction factor}\label{subsec:friction}

The mean streamwise pressure gradient in the range $1000\leq\Rey\leq15.000$ was obtained
by measuring pressure differences between two pressure holes in an angular distance
$\Delta \varphi = \pi/4$. Measurements were only made while the sphere was moving on the
side of the torus opposite of the bore holes. For $\Rey < 18.000$ the pressure difference
$\Delta p = p(\pi/2+\pi/8) - p(\pi/2-\pi/8)$ was found to be constant during a
sufficiently long time span, i.e.\ any pressure fluctuations were sufficiently small to
measure the mean pressure difference $\overline{\Delta p}$ with a maximum r.m.s.\ value
of 0.4\,Pa.

\begin{figure}
  \centerline{\includegraphics[clip,width=0.98\textwidth,angle=0]{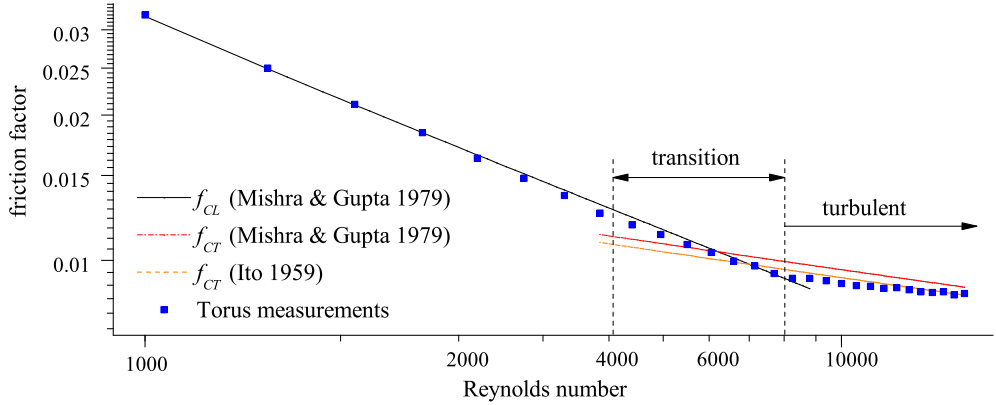}}
  \caption{\label{fig:frictionfactor_logscale}Friction factor measured in the closed torus as a function of the Reynolds number. Straight lines indicate laminar- and turbulent-flow correlations $f_{CL}$ and $f_{CT}$,
  respectively, of \cite{Mishra1979} and the turbulent-flow correlation of \cite{Ito1959}, both for helical pipes.}
\end{figure}

Figure \ref{fig:frictionfactor_logscale} shows the Fanning friction factor
\begin{equation}\label{eq:friction}
    f = \frac {d}{2\rho U^2} \frac {\Delta p}{\Delta \varphi},
\end{equation}
determined from pressure-loss measurements (squares). The measurements are
in good agreement with published data for helical pipes, in particular, for the laminar regime. Deviations
from the laminar friction factor $f_{CL}$ approximated by the power law of
\cite{Mishra1979} become appreciable for $\Rey\gtrsim 2000$ beyond which the measured
friction factor decays more rapidly than the power law. This trend is alleviated by the
onset of oscillations at $\Rey_{c1}$ at which point the slope of the
friction-factor should change. However, we have not enough data to unambiguously resolve
the change of the slope. After a transitional range a further change of the slope of $f$ can be clearly
recognized at $\Rey\approx 8000$. For $\Rey\gtrsim 8000$ the measured data are in
reasonable good agreement with the correlation $f_{CT}$ of \cite{Ito1959} for helical pipes.

\section{Discussion}\label{sec:discussion}

The torus experiment revealed a first instability of the toroidal-pipe flow in the bulk to a
traveling wave at $\Rey=4075$. As the Reynolds number is increased from its critical
value the amplitude of the fundamental Fourier mode grows continuously and homogeneously, indicating a
forward Hopf bifurcation. For $\Rey\gtrsim 4400$ the fundamental frequency develops
sidebands. We anticipate that the sidebands are caused by an incommensurate frequency.
The traveling wave represents a large-scale structure occupying the full cross section of
the pipe. In addition, we find intermittent small-scale and nearly streamwise vortices
which are localized in the wall layer in the vicinity of the impingement point C. For
$\Rey\gtrsim 8000$ the spectrum becomes broadband and the signal exhibits high-frequency
fluctuations. The corresponding threshold Reynolds number bracketed by PIV was confirmed
by pressure-drop measurements which revealed a continuous dependence of the pressure drop
on $\Rey$ with a sudden reduction of the slope at $\Rey\approx 8000$.

Since flow instabilities in toroidal pipes have merely been considered by \cite{Piazza2011}, albeit for periodic boundary conditions, our results can only be further compared with the results for helical pipes of  \cite{Sreenivasan1983} and \cite{Webster1993,Webster1997}.
Regarding the friction factor measurements in the transitional regime we compare our results with those of \cite{Cioncolini2006} for helical pipes. When comparing with helically coiled pipes it is tacitly assumed that the particular driving mechanism in the present experiment does not significantly modify the bulk flow, nor does the small pitch of the helical winding \citep[see e.g.][]{Berger1983,Piazza2011}. Common to all these investigations, the loss
of stability of the steady basic flow occurs in the range $4000\leq\Rey_{c1}\leq6000$
with a subsequent traveling-wave regime for $\Rey>\Rey_{c1}$.

\cite{Sreenivasan1983} used a curvature of $\delta =17.2^{-1}=0.058$ which is
similar to the present value of $\delta=0.049$. Our quantitative results are in good
agreement with their qualitative observations. \cite{Sreenivasan1983} detected periodic
and quasiperiodic low frequency oscillations near the inner wall of the pipe and
high-frequency bursts near its outer wall. The signals were detected by two hot wire
probes in the midplane and at $0.125d$ from the wall at points A and C. The slightly
larger critical Reynolds number of $\Rey=4200$ obtained by \cite{Sreenivasan1983} may be
due to the small difference in $\delta$ and/or the small amplitude of the traveling-wave
at the loci of the hot-wire probes (cf.\ figure \ref{fig:Re4300_Amplitudes}).
Furthermore, the low-frequency signals measured by \cite{Sreenivasan1983} at $\Rey=5000$
have neither a constant amplitude nor a constant frequency. In fact, they are very
similar to our result for the quasiperiodic regime (see e.g.\ $P5$ in figure
\ref{fig:timetraces_3s_5points}). Also the coexistence of a low-frequency modulated wave
with short, high-frequency bursts in different regions of the flow for $\Rey=5870$ is
consistent with the present findings. As \cite{Sreenivasan1983} did neither specify any
wavelengths nor amplitudes a more quantitative comparison cannot be made.

\cite{Webster1993} investigated a nominally fully developed flow through a helically
coiled pipe with a curvature of $\delta = 18.2^{-1}=0.055$ using LDV. The curvature is even closer to the present value. In the range $5060\leq\Rey\leq6330$ the
authors found periodic flow oscillations with a constant fundamental Strouhal number
$\hat{f}=0.25$ measured in the inner half of the pipe cross section. This is
qualitatively compatible with our result. However, we find a nearly linear variation of
the fundamental frequency from $\hat{f}(\Rey=5060)=0.32$ to $\hat{f}(\Rey=6330)=0.15$.

When comparing with the above investigations it must be noted that \cite{Sreenivasan1983} used a long straight-pipe section before the inlet of the coiled pipe, whereas \cite{Webster1993} used a 'flow straightener' in the straight-pipe section, about thirty pipe diameters upstream of the inlet of the curved pipe. For Reynolds numbers at which
the flow is turbulent in the straight pipe two experiments may represent quite different
inlet conditions for the curved pipe. This could be the reason why the first critical
Reynolds number $\Rey_{c1}=5060$ determined by \cite{Webster1993} is well above the
critical Reynolds number determined by \cite{Sreenivasan1983}, and also larger than the
present critical Reynolds number $\Rey_{c1}=4075$.

The numerical simulations of \cite{Webster1997} revealed that the amplitude of the
traveling wave is very small near the midplane and that the maximum streamwise velocity
perturbations arise in the center of region I. This is confirmed by our measurements
(compare figure 7 of \cite{Webster1997} with figures
\ref{fig:Re4300_phaverfluctvel_8subfigs_all} and \ref{fig:Re4300_Amplitudes}).
Wavelengths and phase velocities are of the same order of magnitude, even though
differences remain.

The only study to date of the transition to turbulence in a torus is due to
\cite{Piazza2011}. For curvatures $\delta=0.1$ and $\delta = 0.3$ they numerically
simulated the flow in a torus, driven by an artificial azimuthal body force, for periodic boundary conditions and several
Reynolds numbers in the range from 3500 to 14700. Upon an increase of the Reynolds number
\cite{Piazza2011} found stationary, periodic, quasi-periodic and chaotic flows. The flow
states found are in qualitative agreement with our experimental results. However, there
are differences.

For $\delta=0.3$, a value much larger than the present curvature $\delta=0.049$,
\cite{Piazza2011} reported a supercritical Hopf bifurcation at $\Rey_c=4575$, giving rise
to a traveling wave along the Dean vortices. A second Hopf bifurcation between
$5042<\Rey<5270$ led to a quasi-periodic flow where the additional frequency is
associated with oblique vortices located near the outer equator at C. While the sequence
of bifurcations and the regions where the two waves have a sizable amplitude agree
qualitatively with our results, the symmetry with respect to the midplane differs,
because both modes found by \cite{Piazza2011} are anti-symmetric with respect to the
equatorial plane.

For $\delta=0.1$, a value much closer but yet a factor of two larger than the present
curvature, the periodic and quasi-periodic waves found by \cite{Piazza2011} are
mirror symmetric as in our measurements, but the first instability at $\Rey_c=5175$ is a
saddle-node bifurcation associated with a subcritical Hopf bifurcation. In contrast, we
find a supercritical Hopf bifurcation for $\delta=0.049$. This difference might be
attributed to the different curvatures investigated. It is also not precluded that
the different boundary conditions lead to different finite-amplitude flow patterns when different modes
of the corresponding linear-stability problem are only weakly damped.

\cite{Piazza2011} have sketched a tentative map of flow states in the $(\delta,\Rey)$-plane (their figure 30). Accordingly, the range of chaotic flow is strongly stabilized as $\delta$ is increased from zero, giving way to symmetry-breaking bifurcations at larger $\delta$. In order to elaborate their flow map, further experimental investigations for smaller $\delta$ would be desirable.

Despite of the differences in the setup and the boundary conditions of the above studies on curved pipe flow, there exists no sharp jump in the friction factor upon transition to turbulence as is known from the flow in straight pipes. In the present toroidal pipe flow the flow changes gradually and the friction factor is a monotonous function of $\Rey$. Thus no distinct onset of turbulence can be detected by pressure drop measurements. However, we found indications for an onset of chaotic motion at $\Rey_{c2}\approx 4400$ and a clear change of the slope of $f$ in the logarithmic friction-factor diagram at $\Rey\approx 8000$ which we could correlate, by PIV measurements, to the onset of irregular high-frequency fluctuations.

For helically coiled pipes \cite{Cioncolini2006} analyzed the influence of coil curvature by examination of friction factor profiles. Their findings are in very good qualitative agreement, as they report a \textit{gradual discontinuity} in the friction factor profiles, marking the end of the turbulence emergence process. It must be noted, however, that their results indicate the discontinuity at $\Rey\approx 7300$ for $\delta=0.049$ (according to their equation (13)), which is considerably lower than $\Rey\approx 8000$ found in the present investigation.

\section{Summary}\label{sec:conclusio}

A novel experiment has been set up which realizes a flow with a precisely adjustable flow
rate in a toroidal pipe, allowing to accurately measure the cross-sectional velocity
field using optical techniques. The facility has proven a useful tool to explore the
transition to turbulence in a curved pipe.

Using SPIV distinct flow states have been identified and characterized for the curvature $\delta =0.049$. Mean velocity profiles along the vertical and horizontal axes of
the pipe cross section have been measured as function of the Reynolds number. Moreover,
instantaneous flow fields were obtained for different Reynolds numbers. The first
critical Reynolds number $\Rey_{c1}=4075\pm2\%$ has been determined very accurately. The
bifurcation is of supercritical Hopf type. A further critical Reynolds number was found
at $\Rey_{c2}\approx4400$ at which the flow becomes presumably quasi-periodic. The space
resolved measurements allowed to identify the spatial and temporal structures of the
dominant oscillatory mode just above the threshold $\Rey_{c1}$ as well as large and small scale fluctuations which arise in different regions of the flow for $\Rey>\Rey_{c2}$. The SPIV measurements have shown that the flow becomes turbulent at $\Rey\approx 8000$, a value consistent with the shape of the friction factor curve which has
been measured in the range of $1000\leq\Rey\leq15.000$.

The present investigation, using a single curvature, has confirmed the scenario of
bifurcations that precede the onset of turbulent flow. Perhaps most importantly, we have
measured the traveling wave resulting from the first supercritical instability fully
resolved in the cross-sectional plane and in time. These measurements allowed to identify
the interplay between streamwise vortices and streaks on which the traveling wave is
based. From these observations we deduce that the first instability is caused by the
alternating radial gradients (under $\alpha\approx\pi/4$) of the streamwise velocity near
region I. This result is in some contrast to previous interpretations in terms of a
centrifugal instability.

In view of the sparsity of detailed results on transitional toroidal pipe flow and the
qualitative and quantitative differences among the few previous investigations a more
systematic variation of the curvature would be desirable in order to clarify the
remaining open points. Of particular interest would be the dependence of the first
critical Reynolds number on the curvature, the existence ranges of other flow regimes
which arise through higher-order bifurcations, and the competition with turbulent spots
which may arise on a decrease of the radius ratio. Another interesting aspect would be
the continuation of the unstable wave solutions of the Navier--Stokes equations \citep{Faisst2003,Wedin2004} as the
curvature is increased from zero to curved pipe flow for which stable waves are known to exist.

\bibliographystyle{jfm}

\bibliography{Literatur_JK}

\begin{thebibliography}{34}
\expandafter\ifx\csname natexlab\endcsname\relax\def\natexlab#1{#1}\fi

\bibitem[Adler(1934)]{Adler1934}
{\sc Adler, M.} 1934 Flow in curved tubes. {\em Z. Angew. Math. Mech.\/} {\bf
  14}, 257--275.

\bibitem[Avila {\em et~al.\/}(2011)Avila, Moxey, de~Lozar, Avila, Barkley \&
  Hof]{Avila2011}
{\sc Avila, Kerstin, Moxey, David, de~Lozar, Alberto, Avila, Marc, Barkley,
  Dwight \& Hof, Bj\"{o}rn} 2011 The onset of turbulence in pipe flow. {\em
  Science\/} {\bf 333}~(6039), 192--196.

\bibitem[Berger {\em et~al.\/}(1983)Berger, Talbot \& Yao]{Berger1983}
{\sc Berger, S.~A., Talbot, L. \& Yao, L.~S.} 1983 Flow in curved pipes. {\em
  Annu. Rev. Fluid Mech.\/} {\bf 15}, 461--512.

\bibitem[Budwig(1994)]{Budwig1994}
{\sc Budwig, R.} 1994 Refractive index matching methods for liquid flow
  investigations. {\em Exp. Fluids\/} {\bf 17}, 350--355.

\bibitem[Cioncolini \& Santini(2006)]{Cioncolini2006}
{\sc Cioncolini, Andrea \& Santini, Lorenzo} 2006 An experimental investigation
  regarding the laminar to turbulent flow transition in helically coiled pipes.
  {\em Exp. Therm. Fluid Sci.\/} {\bf 30}~(4), 367 -- 380.

\bibitem[Dean(1927)]{Dean1927}
{\sc Dean, W.~R.} 1927 Note on the motion of fluid in a curved pipe. {\em
  Philos. Mag.\/} {\bf 4}, 208--223.

\bibitem[Dean(1928)]{Dean1928}
{\sc Dean, W.~R.} 1928 The streamline motion of fluid in a curved pipe. {\em
  Philos. Mag.\/} {\bf 7}, 673--695.

\bibitem[van Doorne \& Westerweel(2007)]{Doorne07}
{\sc van Doorne, C. W.~H. \& Westerweel, J.} 2007 Measurement of laminar,
  transitional and turbulent pipe flow using stereoscopic-{PIV}. {\em Exp.
  Fluids\/} {\bf 42}, 259--279.

\bibitem[Drazin \& Reid(1981)]{Drazin1981}
{\sc Drazin, P.~G. \& Reid, W.~H.} 1981 {\em Hydrodynamic Stability\/}.
  Cambridge: Cambridge University Press.

\bibitem[Eckhardt(2008)]{Eckhardt2008}
{\sc Eckhardt, Bruno} 2008 Turbulence transition in pipe flow: some open
  questions. {\em Nonlinearity\/} {\bf 21}~(1), T1.

\bibitem[Eckhardt {\em et~al.\/}(2007)Eckhardt, Schneider, Hof \&
  Westerweel]{Eckhardt2007}
{\sc Eckhardt, B., Schneider, T.~M., Hof, B. \& Westerweel, J.} 2007 Turbulent
  transition in pipe flow. {\em Annu. Rev. Fluid Mech.\/} {\bf 39}, 447--468.

\bibitem[Faisst \& Eckhardt(2003)]{Faisst2003}
{\sc Faisst, H. \& Eckhardt, B.} 2003 Traveling waves in pipe flow. {\em Phys.
  Rev. Lett.\/} {\bf 91}, 224502.

\bibitem[Hasson(1955)]{Hasson1955}
{\sc Hasson, D.} 1955 Streamline flow resistance in coils. {\em Res.
  Corresp.\/} {\bf 1}, 1.

\bibitem[Hewitt {\em et~al.\/}(2011)Hewitt, Hazel, Clarke \&
  Denier]{Hewitt2011}
{\sc Hewitt, R.~E., Hazel, A.~L., Clarke, R.~J. \& Denier, J.~P.} 2011 Unsteady
  flow in a rotating torus after a sudden change in rotation rate. {\em J.
  Fluid Mech.\/} {\bf 688}, 88--119.

\bibitem[Hof {\em et~al.\/}(2004)Hof, van Doorne, Westerweel, Nieuwstadt \&
  Faisst]{Hof2004}
{\sc Hof, B., van Doorne, C. W.~H., Westerweel, J., Nieuwstadt, F. T.~M. \&
  Faisst, H.} 2004 Experimental observation of nonlinear traveling waves in
  turbulent pipe flow. {\em Science\/} {\bf 305}, 1594--1598.

\bibitem[Hopkins {\em et~al.\/}(2000)Hopkins, Kelly, Wexler \&
  Prasad]{Hopkins2000}
{\sc Hopkins, L.~M., Kelly, J.~T., Wexler, A.~S. \& Prasad, A.~K.} 2000
  Particle image velocimetry measurements in complex geometries. {\em Exp.
  Fluids\/} {\bf 29}, 91--95.

\bibitem[H\"{u}ttl \& Friedrich(2000)]{Huttl2000}
{\sc H\"{u}ttl, T.~J. \& Friedrich, R.} 2000 Influence of curvature and torsion
  on turbulent flow in helically coiled pipes. {\em Int. J. Heat Fluid Flow\/}
  {\bf 21}, 345--353.

\bibitem[H\"{u}ttl \& Friedrich(2001)]{Huttl2001}
{\sc H\"{u}ttl, T.~J. \& Friedrich, R.} 2001 Direct numerical simulation of
  turbulent flows in curved and helically coiled pipes. {\em Comput. Fluids\/}
  {\bf 30}, 591--605.

\bibitem[Ito(1959)]{Ito1959}
{\sc Ito, H.} 1959 Friction factors for turbulent flow in curved pipes. {\em J.
  Basic Eng. Trans. ASME\/} {\bf 81}, 123--134.

\bibitem[Lowe \& Kutt(1992)]{Lowe1992}
{\sc Lowe, M.~L. \& Kutt, P.~H.} 1992 Refraction through cylindrical tubes.
  {\em Exp. Fluids\/} {\bf 13}, 315--320.

\bibitem[Madden \& Mullin(1994)]{Madden1994}
{\sc Madden, F.~N. \& Mullin, T.} 1994 The spin-up from rest of a fluid-filled
  torus. {\em J. Fluid Mech.\/} {\bf 265}, 217--244.

\bibitem[Mishra \& Gupta(1979)]{Mishra1979}
{\sc Mishra, P. \& Gupta, S.~N.} 1979 Momentum transfer in curved pipes 1.
  newtonian fluids; 2. non-newtonian fluids. {\em Ind. Eng. Chem. Process Des.
  Dev.\/} {\bf 18}, 130.

\bibitem[Mullin(2011)]{Mullin2011}
{\sc Mullin, T.} 2011 Experimental studies of transition to turbulence in a
  pipe. {\em Annu. Rev. Fluid Mech.\/} {\bf 43}~(1), 1--24.

\bibitem[Naphon \& Wongwises(2006)]{Naphon2006}
{\sc Naphon, P. \& Wongwises, S.} 2006 A review of flow and heat transfer
  characteristics in curved tubes. {\em Renewable Sustainable Energy Rev.\/}
  {\bf 10}, 463--490.

\bibitem[Piazza \& Ciofalo(2011)]{Piazza2011}
{\sc Piazza, Ivan~Di \& Ciofalo, Michele} 2011 Transition to turbulence in
  toroidal pipes. {\em J. Fluid Mech.\/} {\bf 687}, 72--117.

\bibitem[del Pino {\em et~al.\/}(2008)del Pino, Hewitt, Clarke, Mullin \&
  Denier]{Pino2008}
{\sc del Pino, C., Hewitt, R.~E., Clarke, R.~J., Mullin, T. \& Denier, J.~P.}
  2008 Unsteady fronts in the spin-down of a fluid-filled torus. {\em Phys.
  Fluids\/} {\bf 20}~(12), 124104.

\bibitem[Reynolds(1883)]{Reynolds1883}
{\sc Reynolds, O.} 1883 An experimental investigation of the circumstances
  which determine whether the motion of water shall be direct or sinuous and of
  the law of resistance in parallel channels. {\em Philos. Trans. R. Soc. Lond.
  Ser. A\/} {\bf 174}, 935.

\bibitem[Sreenivasan \& Strykowski(1983)]{Sreenivasan1983}
{\sc Sreenivasan, K.~R. \& Strykowski, P.~J.} 1983 Stabilization effects in
  flow through helically coiled pipes. {\em Exp. Fluids\/} {\bf 1}, 31--36.

\bibitem[Taylor(1929)]{Taylor1929}
{\sc Taylor, G.~I.} 1929 The criterion for turbulence in curved pipes. {\em
  Proc. Roy. Soc. London A\/} {\bf 124}, 243--249.

\bibitem[Vashisth {\em et~al.\/}(2008)Vashisth, Kumar \& Nigam]{Vashisth2008}
{\sc Vashisth, Subhashini, Kumar, Vimal \& Nigam, Krishna D.~P.} 2008 A review
  on the potential applications of curved geometries in process industry. {\em
  Industrial \& Engineering Chemistry Research\/} {\bf 47}~(10), 3291--3337.

\bibitem[Webster \& Humphrey(1993)]{Webster1993}
{\sc Webster, D.~R. \& Humphrey, J. A.~C.} 1993 Experimental observation of
  flow instability in a helical coil. {\em Journal of Fluids Engineering\/}
  {\bf 115}~(3), 436--443.

\bibitem[Webster \& Humphrey(1997)]{Webster1997}
{\sc Webster, D.~R. \& Humphrey, J. A.~C.} 1997 Traveling wave instability in
  helical coil flow. {\em Phys. Fluids\/} {\bf 9}, 407--418.

\bibitem[Wedin \& Kerswell(2004)]{Wedin2004}
{\sc Wedin, H. \& Kerswell, R.~R.} 2004 Exact coherent structures in pipe flow:
  traveling wave solutions. {\em J. Fluid Mech.\/} {\bf 508}, 333--371.

\bibitem[White(1929)]{White1929}
{\sc White, C.~M.} 1929 Streamline flow through curved pipes. {\em Proc. Roy.
  Soc. London A\/} {\bf 123}, 645--663.

\end{thebibliography}

\end{document}